\documentclass[a4paper]{article}

\usepackage[widelayout,hyperrefblack]{research17}
\usepackage{threeparttable}
\usepackage{diagbox}
\usepackage{array}
\usepackage{boldline}
\usepackage{bbm}
\usepackage{algorithmicx}
\usepackage{algpseudocode}
\usepackage{subfigure}
\usepackage{appendix}

\allowdisplaybreaks
\usetikzlibrary{datavisualization} 

\usepackage{pgfplots}
\usepgfplotslibrary{units}
\usetikzlibrary{spy,backgrounds}
\usetikzlibrary{decorations.markings}
\usepackage{pgfplotstable}

\renewcommand{\indicatorfunction}{\mathop{}\!\mathbbm{1}}

\renewcommand{\indicatorfunction}{\mathop{}\!\mathbbm{1}}

\newcommand{\llg}[1]{{\color{black}#1}}

\makeatletter
\def\thanks#1{\protected@xdef\@thanks{\@thanks
		\protect\footnotetext{#1}}}
\makeatother

\begin{document}

\title{On the Low-SNR Asymptotic Capacity of Two Types of Optical Wireless Channels under Average-Intensity Constraints}

\author{ Longguang Li
	\thanks{This work is supported by the National Natural Science Foundation of China under Grant No. 62071489. }
\thanks{L. Li is with SJTU Paris Elite Institute of Technology, Shanghai Jiao Tong University, Shanghai, China  (e-mail: llg9012@sjtu.edu.cn).} 
%	\thanks{R. H. Chen is with National Digital Switching System Engineering and Technological Research Center, Zhengzhou, China (e-mail: RyanChen1210@163.com).} 
%\thanks{J. Zhou is with Shaoxing University, Shaoxing, China. He was with the CAS Key Laboratory of Wireless-Optical Communications, University of Science and Technology of China, Hefei, China (e-mail: jzee@ustc.edu.cn).}
}

\maketitle

\begin{abstract}
In this paper, we study two types of optical wireless channels under average-intensity constraints. One is called the Gaussian optical intensity channel, where the channel output models the converted electrical current corrupted by \llg{additive white Gaussian noise}. The other one is the Poisson optical intensity channel, where the channel output models the number of received photons whose arrival rates are corrupted by \llg{a dark current}. When the average input intensity $\mathcal{E}$ is small, the capacity of the Gaussian optical intensity channel is shown to scale as $\mathcal{E}\sqrt{\frac{\log\frac{1}{\mathcal{E}}}{2}}$, and the capacity of the Poisson optical intensity channel as $\mathcal{E}\log\log\frac{1}{\mathcal{E}}$. \llg{This closes the existing capacity gaps in these two types of channels.}
%This paper investigates the sum capacity of two-user optical intensity multiple access channels
%with per-user peak- or/and average-intensity constraints. By leveraging tools on decompositions of several continuous and discrete random variables, we derive several bounds on the sum capacity. In the high signal-to-noise ratio (SNR) regime, some bounds asymptotically match or approach to the sum capacity, thus complementing or reducing the existing gaps on high-SNR asymptotic sum capacity. At moderate SNR, some bounds are also fairly close to the sum capacity.
\end{abstract}

\section{Introduction}
Intensity-modulation and direct-detection (IM-DD) is widely \llg{adopted} in most current optical wireless communication (OWC) systems because of its \llg{low cost} and convenient implementation. In this scheme, information is carried on  the modulated intensity of the optical light, \llg{and detected via a photodetector measuring the incoming intensity}~\cite{gagliardikarp76_1, coxackermanhelkeybetts97_1, leerandaelbreyerkoonen09_1}.  The optical signal in this scheme is real and nonnegative, which \llg{leads to fundamental differences to traditional radio-frequency communication. There exist several different IM-DD based channel models~\cite{kahnbarry97_1,bradyverdu90_1,faridhranilovic09_1,lapidothmoserwigger09_7,moser12_3}, whose exact capacity characterizations are still open} problems~\cite{smith71_1,mmmovickschischang05_1,sharma2010transition}. In the existing literature,  progress has been made \llg{in two aspects}. One aspect is on deriving capacity bounds or characterizing asymptotic capacity in the high or low signal-to-noise ratio (SNR) regime. Capacity upper and lower bounds and high/low-SNR asymptotic results on single-input single-output channels are derived~\cite{bradyverdu90_1,lapidothmoserwigger09_7,moser12_3,faridhranilovic09_1,faridhranilovic10_1,cheraghchi2021non,cheraghchi2019improved}, and similar results are extended to general multiple-input single- and multiple-output channels~\cite{limoserwangwigger20_1,chen2021MISO,chaaban2020capacity}. The other aspect is on characterizing properties of the capacity-achieving input distribution, e.g., the discreteness and finiteness of its support~\cite{mmmovickschischang05_1,
smith71_1,sharma2010transition,cao2013capacity,cao2013capacity_2}. Recently, bounds on the cardinality of its support \llg{were shown in} \llg{\cite{dytso2019capacity,dytso2021properties}}. 

This paper studies two types of OWC channels \llg{under average-intensity constraints}, and focuses on characterizing the low-SNR asymptotic capacity. The first considered channel is the Gaussian optical intensity channel. In the related existing work~\cite{lapidothmoserwigger09_7},  when the peak or both the peak and average intensity of the input are limited, the low-SNR asymptotic capacity \llg{is characterized exactly, and is expressed in terms} of the maximum variance among all admissible input distributions. In the case where \llg{only} the average intensity of the input is limited, the existing result is \llg{restricted} to the scaling order of the low-SNR asymptotic capacity. \llg{The difficulty comes from the fact that the low-SNR capacity} is no longer captured by the maximum variance of the input distributions, which can be arbitrarily large. Specifically, when the average intensity of the input is limited to be no larger than $\mathcal{E}$, the low-SNR capacity shown in~\cite{lapidothmoserwigger09_7} scales as $a_{\text{G}}\mathcal{E}\sqrt{ {\log \frac{1}{\mathcal{E}}} }$, where constant $a_{\text{G}}$ satisfies $\frac{1}{\sqrt{2}}\leq a_{\text{G}} \leq 2$.

The other considered channel is the Poisson optical intensity channel. When there is positive dark current in the channel,~\cite{lapidothshapirovenkatesanwang11_1} shows similar low-SNR capacity results as in the Gaussian optical intensity channel. When the average intensity of the input is limited to be no larger than $\mathcal{E}$, the low-SNR capacity scales as $a_{\text{P}}\mathcal{E}{ {\log\log \frac{1}{\mathcal{E}}} }$, where constant $a_{\text{P}}$ satisfies $\frac{1}{{2}}\leq a_{\text{P}} \leq 2$. 

 In this paper, we show that $a_\text{G}=\frac{1}{\sqrt{2}}$ and  $a_\text{P}=1$. Hence, the low-SNR asymptotic capacity of the Gaussian and Poisson optical intensity channels scale exactly as $\mathcal{E}\sqrt{ \frac{\log \frac{1}{\mathcal{E}}}{2} }$ and $\mathcal{E}\log\log\frac{1}{\mathcal{E}}$, respectively. \llg{The results are proved using a duality-based upper bound to capacity that relies on a carefully chosen auxiliary distribution, and the achievability part leverages} tools from the data processing inequality, Fano's inequality, and the maximum a posteriori probability (MAP) decision rule.

\llg{The paper is organized as follows. We end the introduction with a few notation's conventions. Section~\ref{sec:channel-model} describes in detail the two investigated channel models. In Section~\ref{eq:mainresults}, we present the low-SNR asymptotic capacity results and the corresponding proofs of the inverstigated channels. We will conclude in Section~\ref{conclu}.}

\llg{
\textbf{Notation:} We use uppercase letters for random variables, e.g., $X$, and for their realizations lowercase letters, e.g., $x$. Entropy of a random variable is denoted by $\HH(\cdot)$, and mutual information by $\II(\cdot ; \cdot)$. The expectation of a random variable is denoted by $\E{\cdot}$. We use $\const{D}(\cdot\|\cdot)$ to denote the Kullback--Leibler divergence, and $\lfloor a \rfloor$ to denote the largest integer not exceeding $a$. We denote $ \phi(x) \overset{\text{def}}{=} \frac{1}{\sqrt{2\pi}} e^{-\frac{x^2}{2}}$, and $\mathcal{Q}(x) \overset{\text{def}}{=} \int_{x}^{\infty}\phi(t)\text{d}t$. $\log (\cdot)$ denotes the logarithm to the base of $e$. We use $f(x)~\dot{=}~g(x)$ to indicate functions $f(x)$ and $g(x)$ \llg{satisfying} $\lim_{x\rightarrow 0^+} \frac{f(x)}{g(x)}=1$, and $f(x)~\dot{\leq}~g(x)$ and  $f(x)~\dot{\geq}~g(x)$ are defined similarly.
}
%, vectors and matrix are boldfaced as $\rv{h}$, $\mat{X}_j$ and $\mat{X}$, while their realizations are typeset as $h$, $\matt{X}_j$ and $\matt{X}$, respectively. Differential entropy is denoted by $\hh(\cdot)$ and mutual information by $\II(\cdot ; \cdot)$. The expectation and variance of a random variable is denoted by $\E{\cdot}$ and $\Var{\cdot}$ respectively. $\|\cdot\|_{1}$ and $\|\cdot\|_\mathsf{F}$ denote the $\mathcal{L}_{1}$- and Frobenius-norm respectively. $\log (\cdot)$ and $\ln(\cdot)$ denote the logarithm to the base of $2$ or $e$ respectively.  Sets are typeset in a special font (such as $\set{A}$), and $\set{R}^{m \times n}$ ($\set{R}_+^{m \times n}$) denotes real-valued (and nonnegative) set. Symbol $\doteq$ ($\dot{\geqslant}$ or $\dot{\leqslant}$) denotes \emph{exponential} equality (inequality)\footnote{$f(x)\doteq x^b$ denotes $\lim _{x \rightarrow \infty} \frac{\log f(x)}{\log x}=b$.}.
%To simplify the notation, 

\section{Channel Model}
\label{sec:channel-model}
%\subsection{Gaussian Optical Intensity Channel}
The channel output of a Gaussian optical intensity channel is given by
\begin{IEEEeqnarray}{rCl}
Y = x+Z,
\label{eq:singlechannelmodel}
\end{IEEEeqnarray}
where $x$ denotes the channel input, and  $Z$ denotes the Gaussian noise with variance $1$, i.e., 
\begin{IEEEeqnarray}{c}
Z\sim \mathcal{N}(0,1),
\label{gaussiannoise}
\end{IEEEeqnarray}
\llg{independent of the input $X$.}

\llg{Since $x$ is proportional to the optical intensity, it cannot be negative}
\begin{IEEEeqnarray}{c}
x \in \mathbb{R}^+.
\label{eq:mm1}
\end{IEEEeqnarray}
\llg{Considering eye safety and energy consumption, the input must be constrained}
%peak-intensity constraint:
%\begin{IEEEeqnarray}{c}
%\textnormal{Pr}( X \leq 1 ) = 1,
%\label{eq:mm2}
%\end{IEEEeqnarray}
%or/and subject to 
\begin{IEEEeqnarray}{c}
\E{X}  \leq \mathcal{E},
\label{eq:mm3}
\end{IEEEeqnarray}
where $\mathcal{E} >0$ is a given constant.
%where constant $\alpha \in \left(0,\frac{1}{2}\right]$, denotes the average intensity constraint. 

%\subsection{Poisson Optical Intensity Channel}
\llg{The Poisson optical intensity channel with dark current $\lambda > 0$ has a channel output Y that follows}
%Now we consider a Poisson optical intensity channel with dark current $\lambda > 0$.Conditional on the input $x$, the channel output $Y$ follows a Poisson distribution with arrival rate $\lambda+x$, i.e., the channel law $W(\cdot|X=x)$ is
\begin{IEEEeqnarray}{rCl}
W(Y=y|X=x)= e^{-(\lambda+x)}\frac{(\lambda+x)^y}{y!}, \quad y \in \mathbb{N}.
\label{eq:singlechannelmodel2}
\end{IEEEeqnarray}
Input $x$ of this channel also needs to satisfy the constraints~\eqref{eq:mm1} and~\eqref{eq:mm3}. To simplify the notation in the paper, we also \llg{denote} above distribution~\eqref{eq:singlechannelmodel2} as $\text{Poi}_{\lambda+x}(y)$. 
%where constant $\alpha \in \left(0,\frac{1}{2}\right]$, denotes the average intensity constraint. 

The single-letter capacity expression of the channel~\eqref{eq:singlechannelmodel} or~\eqref{eq:singlechannelmodel2} is given by
\begin{flalign}
	\mathsf{C} (\mathcal{E}) = \sup_{p_X \textnormal{ satisfying }  \eqref{eq:mm1} \text{ and } \eqref{eq:mm3}} \II\left(X;Y\right),
\end{flalign}
\llg{where the supremum is over all input distributions satisfying the intensity constraints \eqref{eq:mm1} and \eqref{eq:mm3}}. In the rest of the paper, we use $\mathsf{C}_{\textnormal{G}} (\mathcal{E})$ and $\mathsf{C}_{\textnormal{P}} (\mathcal{E})$ to denote the capacity of Gaussian and Poisson optical intensity channels, respectively.  
\section{Main Result}
\label{eq:mainresults}
\subsection{Gaussian Optical Intensity Channel}
\begin{theorem}
\label{thm1}
The capacity of channel~\eqref{eq:singlechannelmodel} satisfies \llg{
\begin{IEEEeqnarray}{rCl}
\lim_{\mathcal{E} \to 0^{+}} 
\frac{\mathsf{C}_{\textnormal{G}}(\mathcal{E})}{\mathcal{E}\sqrt{ {\log \frac{1}{\mathcal{E}}} }} = \frac{1}{\sqrt{2}}.
\end{IEEEeqnarray} 
}
\end{theorem}
%\begin{IEEEproof}
\llg{We prove Theorem~\ref{thm1} in two steps. Note that it is equivalent to prove 
\begin{IEEEeqnarray}{rCl}
\limsup_{\mathcal{E} \to 0^{+}} 
\frac{\mathsf{C}_{\textnormal{G}}(\mathcal{E})}{\mathcal{E}\sqrt{ {\log \frac{1}{\mathcal{E}}} }} \leq \frac{1}{\sqrt{2}},
\label{eq:11}
\end{IEEEeqnarray}
and 
\begin{IEEEeqnarray}{rCl}
%\liminf_{\mathcal{E} \to 0^{+}^+} 
\liminf_{\mathcal{E} \to 0^{+}} 
\frac{\mathsf{C}_{\textnormal{G}}(\mathcal{E}) }{\mathcal{E}\sqrt{ {\log \frac{1}{\mathcal{E}}} }} \geq \frac{1}{\sqrt{2}}.
\label{eq:12}
\end{IEEEeqnarray}
We will prove Eq.~\eqref{eq:11} in Section~\ref{secteq11}, and prove Eq.~\eqref{eq:12} in Section~\ref{secteq12}. }
\subsubsection{Proof of Eq.~\eqref{eq:11}} 
\label{secteq11}
\llg{We first present a lemma that is useful here. }
%We first present a lemma that will be useful in the proof here.
\llg{
\begin{lemma}
\label{lemm2}
Fix a real number $\xi > 0$. Then, for any $\tau \geq 0$, 
\begin{IEEEeqnarray}{rCl}
\phi(\xi-\tau) \leq \phi (\xi)+\frac{2\tau}{\xi}.
\label{eq:mg}
\end{IEEEeqnarray}
\end{lemma}
\begin{IEEEproof}
See Appendix~\ref{lem:phifunc}.
\end{IEEEproof}
}
\llg{Now we prove~\eqref{eq:11}. Capacity can be upper-bounded using the following bound based on duality~\cite{moser04_1}:}
\begin{IEEEeqnarray}{rCl}
  \label{eq:E101}
  \mathsf{C}_{\textnormal{G}} &\leq& \sup_{p_X} \E{\const{D}\bigl(W(\cdot|X)\big\|R(\cdot)\bigr)}
\label{eq:dualbndorg} \\
 &=& \sup_{p_X} \E{ -\int_{-\infty}^{\infty} W(y|X)\log R(y) \dd y } -\frac{1}{2}\log 2\pi e,
\label{eq:dualbnd}
\end{IEEEeqnarray}
where $W(\cdot|x)$ denotes the conditional output distribution given the input $X=x$, and $R(\cdot)$ \llg{denotes an arbitrary distribution on} the output space. More details on the duality capacity bound can be found in~\cite{moser04_1}. 

 When $\mathcal{E}$ is sufficiently small, let $t=a_{\textnormal{G}}\sqrt{\log \frac{1}{\mathcal{E}}}$ with $a_{\textnormal{G}}>\sqrt{2}$, and $\beta=e^{-\frac{t^2}{2}}$. \llg{Note that} $\beta \in (0,1)$. We choose the auxiliary distribution $R(\cdot)$ as\footnote{ \llg{Note that $\int_{-\infty}^{\infty}R(y)\dd y = \int_{-\infty}^{t}R(y)\dd y + \int_{t}^{\infty}R(y)\dd y =  \frac{1-\beta}{\mathcal{Q}(-t)}\int_{-\infty}^{t}\frac{e^{-\frac{t^2}{2}}}{\sqrt{2\pi}}\dd y + \beta\int_{t}^{\infty}e^{-(y-t)}\dd y = \frac{1-\beta}{\mathcal{Q}(-t)} \cdot \mathcal{Q}(-t)+\beta \cdot 1=1$, which verifies that $R(\cdot)$ is indeed a distribution.} } 
\begin{IEEEeqnarray}{c}
  R(y) =
  \begin{cases} 
     \frac{1-\beta}{\sqrt{2\pi}\mathcal{Q}(-t)}e^{-\frac{y^2}{2}} & \textnormal{if } y \leq t,
    \\
    \beta e^{-(y-t)} & \textnormal{otherwise}.
  \end{cases}
\end{IEEEeqnarray}
%where the parameter $\beta\in(0,1)$ will both be specified later.

The expectation term at the RHS of~\eqref{eq:dualbnd} can be expanded as
\begin{IEEEeqnarray}{rCl}
  \IEEEeqnarraymulticol{3}{l}{%
    \E{-\int_{-\infty}^{\infty} W(y|X)\log R(y) \dd y }
  }\nonumber\\*\quad%
&=&  \E{-\int_{-\infty}^{t} W(y|X)\log R(y) \dd y }+ \E{ -\int_{t}^{\infty} {W(y|X)\log{{R(y)} }} \dd y }. 
\label{eq:firstsecond}
\end{IEEEeqnarray}

For the first term at the RHS of~\eqref{eq:firstsecond}, we have \llg{
\begin{IEEEeqnarray}{rCl}
  \IEEEeqnarraymulticol{3}{l}{%
    \E{-\int_{-\infty}^{t} W(y|X)\log R(y) \dd y }
  }\nonumber\\*\quad%
  & = & \E{ -\int_{-\infty}^{t}
  \frac{1}{\sqrt{2\pi}} \ope^{-\frac{(y-X)^2}{2}}
  \left(\log{\frac{1-\beta}{\sqrt{2\pi}\mathcal{Q}(-t)}} - \frac{y^2}{2}\right) \dd y }
  \\ 
  & = & \mathsf{E}\bigg[ - \log{\frac{1-\beta}{\sqrt{2\pi}\mathcal{Q}(-t)}} \underbrace{ \int_{-\infty}^{t}
  \frac{1}{\sqrt{2\pi}} \ope^{-\frac{(y-X)^2}{2}}
    \dd y}_{=\mathcal{Q}(X-t)}   + \frac{1}{2} \underbrace{ \int_{-\infty}^{t}
  \frac{y^2}{\sqrt{2\pi}} \ope^{-\frac{(y-X)^2}{2}} \dd y }_{(1+X^2)\mathcal{Q}(X-t)-(X+t)\phi(X-t)}  \bigg]
  \\ 
  & = & \mathsf{E}\bigg[-\Q(X-t)\log \frac{1-\beta}{\sqrt{2\pi}\mathcal{Q}(-t)} 
  + \frac{1}{2}\Q(X-t)  \nonumber\\
&&\hspace{0.5cm}+ \frac{X^2}{2} \Q(X-t) - \underbrace{\frac{X+t}{2}\phi(X-t)}_{\geq 0}  \bigg]
  \\
  & \leq &\E{ -\Q(X-t)\log \frac{1-\beta}{\sqrt{2\pi}\mathcal{Q}(-t)} 
  + \frac{1}{2}\Q(X-t) + \frac{X^2}{2} \Q(X-t) }
  \label{eq:E207} \\
&=& \mathsf{E}\bigg[ \underbrace{\Q(X-t)}_{\leq 1}\log \frac{\sqrt{2\pi e}}{1-\beta}  -\underbrace{\Q(X-t)\log\frac{1}{\mathcal{Q}(-t)}}_{\geq 0}
  + \frac{X^2}{2} \Q(X-t) \bigg] \\
&\leq&   \log \frac{\sqrt{2\pi e}}{1-\beta}
  +  \frac{1}{2}\E{X^2\mathcal{Q}(X-t)} \\
&=&   \frac{1}{2}\log 2\pi e+\log \left(1+\frac{\beta}{1-\beta} \right)
  +  \frac{1}{2}\E{X^2\mathcal{Q}(X-t)}
 \\
&\leq& \frac{1}{2}\log 2\pi e + \frac{\beta}{1-\beta}+  \frac{1}{2}\E{X^2\mathcal{Q}(X-t)} \label{eq:EE207} \\
%&\dot{=}& \frac{1}{2}\log 2\pi e + \mathcal{E}^{\frac{a^2}{2}}+  \frac{1}{2}\E{X^2\mathcal{Q}(X-t)}.
%&\leq&  {-\log \frac{1-\beta}{\sqrt{2\pi e}\mathcal{Q}(-t)}}
%  + \frac{\mathcal{E}(1+t)}{2},
&=& \frac{1}{2}\log 2\pi e + \frac{e^{-\frac{t^2}{2}}}{1-e^{-\frac{t^2}{2}}}+  \frac{1}{2}\E{X^2\mathcal{Q}(X-t)}, \label{eq:EE2077}
\end{IEEEeqnarray}
where~\eqref{eq:EE207} follows from $\log(1+x)\leq x,~x\geq0$ .}

For the second term at the RHS of~\eqref{eq:firstsecond}, we have 
\begin{IEEEeqnarray}{rCl}
  \IEEEeqnarraymulticol{3}{l}{%
\E{ -\int_{t}^{\infty} {W(y|X)\log{{R(y)} }} \dd y }
  }\nonumber\\*\quad%
  & = & \E{ -\int_{t}^{\infty} {\frac{1}{\sqrt{2\pi}}\ope^{-\frac{(y-X)^2}{2}}}
  \left(\log{{\beta}}
    -({y-t}) \right) \dd y }
  \\
  & = & \mathsf{E}\bigg[ -\log{{\beta}} \underbrace{\int_{t}^{\infty} {\frac{1}{\sqrt{2\pi}}\ope^{-\frac{(y-X)^2}{2}}} \dd y}_{\mathcal{Q}(t-X)}
  +  \underbrace{ \int_{t}^{\infty} {\frac{y-t}{\sqrt{2\pi}}\ope^{-\frac{(y-X)^2}{2}}}\dd y }_{({X-t})\mathcal{Q}(t-X)+{\phi(t-X)}}
    \bigg]
  \\
  & = & \mathsf{E}\bigg[-\Q(t-X) \log
  {{\beta}} 
  + ({X-t})\mathcal{Q}(t-X)+{\phi(t-X)}\bigg]   \\
  & = & \mathsf{E} \bigg[ -\Q(t-X) \log
  {{\beta}} 
   + {X}\underbrace{\mathcal{Q}(t-X)}_{\leq 1}-\underbrace{t\mathcal{Q}(t-X)}_{\geq0}+{\phi(t-X)} \bigg]  \\
  & \leq & \E{ \frac{t^2}{2}
  \Q(t-X) + X+{\phi(t-X)} } \\
  & \leq & \E{ \frac{t^2}{2}
  \Q(t-X) + X+{\phi(t)+\frac{2X}{t}} } \label{eq:phiexp} \\
  & \leq & {\phi(t)+\left(1+\frac{2}{t}\right)\mathcal{E}}  + \frac{1}{2}\E{ {t^2}
  \Q(t-X)},
%  & \leq &
%\E{ \log {\frac{\mu}{\beta}}\cdot\Q(t-x) +\> \frac{x}{\mu}+\frac{\phi(t)+\frac{x}{\sqrt{2\pi}}}{{\mu}}  } \\
%& \leq &\E{ \log {\frac{\mu}{\beta}}\cdot\Q(t-x)} + \frac{2\mathcal{E}}{\mu}+\frac{\phi(t)}{\mu}.
%  & = &
% \log {\frac{\mu}{\beta}}\cdot (1-\Q(x-t)) +\> \frac{x}{\mu}+\frac{1}{\sqrt{2\pi}\mu}.
\label{eq:E2210} 
\end{IEEEeqnarray}
where~\eqref{eq:phiexp} follows Lemma~\ref{lemm2}, and~\eqref{eq:E2210} by~\eqref{eq:mm3}.

Combining~\eqref{eq:EE2077} and~\eqref{eq:E2210}, we have
\begin{IEEEeqnarray}{rCl}
  \IEEEeqnarraymulticol{3}{l}{%
    \E{-\int_{-\infty}^{\infty} W(y|X)\log R(y) \dd y }
  }\nonumber\\*\quad%
&\leq&  \frac{1}{2}\log 2\pi e +  \frac{e^{-\frac{t^2}{2}}}{1-e^{-\frac{t^2}{2}}}+{\phi(t)+\left(1+\frac{2}{t}\right)\mathcal{E}}+\frac{1}{2}\E{X^2\mathcal{Q}(X-t)+t^2\mathcal{Q}(t-X)}.
\IEEEeqnarraynumspace 
\label{eq:EE2112}
\end{IEEEeqnarray}

Now, we bound the expectation term at the RHS of~\eqref{eq:EE2112}. By the law of total expectation,
\begin{IEEEeqnarray}{rCl}
\E{X^2\mathcal{Q}(X-t)+t^2\mathcal{Q}(t-X)}
 &=& \Econd{X^2\mathcal{Q}(X-t)+t^2\mathcal{Q}(t-X)}{X\leq t}\text{Pr}(X\leq t) \nonumber\\
&&\>+ \Econd{X^2\mathcal{Q}(X-t)+t^2\mathcal{Q}(t-X)}{X > t}\text{Pr}(X > t). 
\IEEEeqnarraynumspace
\label{eq:EE2113}
% &=& \Econd{X^2+(t^2-X^2)\mathcal{Q}(t-X)}{X\leq t}\text{Pr}(X\leq t) + \Econd{t^2+(X^2-t^2)\mathcal{Q}(X-t)}{X > t}\text{Pr}(X > t) \\
%&\leq& \Econd{Xt+(t+X)(t-X)\mathcal{Q}(t-X)}{X\leq t}\text{Pr}(X\leq t) + \Econd{Xt+(X+t)(X-t)\mathcal{Q}(X-t)}{X > t}\text{Pr}(X > t) \nonumber\\ 
%\label{eq:EE2110}
%\\
%&\leq& \Econd{Xt+(t+X)\phi(t-X)}{X\leq t}\text{Pr}(X\leq t) + \Econd{Xt+(X+t)\phi(X-t)}{X > t}\text{Pr}(X > t) \\
%&=& t\E{X}+ \E{(t+X)\phi(t-X)} \\
%&\leq& \mathcal{E}t+ t\E{\phi(t-X)} + \mathcal{E} \\
%&\leq& \mathcal{E}t+ t\E{\phi(t)+\frac{2X}{t}} + \mathcal{E} \\
%&\leq& \mathcal{E}t+ t\phi(t) + 3\mathcal{E}.
\end{IEEEeqnarray}
The first conditional expectation terms at the RHS of~\eqref{eq:EE2113} can be bounded as 
\begin{IEEEeqnarray}{rCl}
\Econd{X^2\mathcal{Q}(X-t)+t^2\mathcal{Q}(t-X)}{X\leq t} &=&\Econd{X^2(1-\mathcal{Q}(t-X))+t^2\mathcal{Q}(t-X)}{X\leq t}
\IEEEeqnarraynumspace \\
&=& \mathsf{E}\bigg[\underbrace{X^2}_{\leq Xt}+(t^2-X^2)\mathcal{Q}(t-X) \bigg| X\leq t \bigg] \\
 &\leq& \mathsf{E}\bigg[ Xt+(t+X)\underbrace{(t-X)\mathcal{Q}(t-X)}_{\leq \phi(t-X)} \bigg] \bigg| X\leq t \bigg] 
\\
 &\leq& \Econd{Xt+(t+X)\phi(t-X)}{X\leq t} \label{eq:expqfun}\\
 &=& t\Econd{X}{X\leq t} +  \Econd{(t+X)\phi(t-X)}{X\leq t}, 
\label{eq:EEE201}
\end{IEEEeqnarray}
where~\eqref{eq:expqfun} follows by $x\mathcal{Q}(x)\leq \phi (x), \,\, \forall x \geq 0$~\cite{gordon41_1}. Similarly, for the second conditional expectation terms at the RHS of~\eqref{eq:EE2113},
\begin{IEEEeqnarray}{rCl}
\Econd{X^2\mathcal{Q}(X-t)+t^2\mathcal{Q}(t-X)}{X > t} &=& \Econd{t^2+(X^2-t^2)\mathcal{Q}(X-t)}{X > t} \\
 &\leq& \Econd{Xt+(X+t)(X-t)\mathcal{Q}(X-t)}{X > t} 
\\
 &\leq& \Econd{Xt+(X+t)\phi(X-t)}{X > t} \\
 &=& t\Econd{X}{X>t} +  \Econd{(t+X)\phi(X-t)}{X>t}.
\IEEEeqnarraynumspace
\label{eq:EEE200}
\end{IEEEeqnarray}
Substituting~\eqref{eq:EEE200} and~\eqref{eq:EEE201} into~\eqref{eq:EE2113}, we obtain
\begin{IEEEeqnarray}{rCl}
\E{X^2\mathcal{Q}(X-t)+t^2\mathcal{Q}(t-X)}
 &\leq& t\E{{X}}+ \E{(t+X)\phi(t-X)} \\
&=& t\E{{X}}+ \mathsf{E}\bigg[ t\phi(t-X)+X\underbrace{\phi(t-X)}_{\leq \frac{1}{\sqrt{2\pi}}} \bigg]\\
 &\leq& t\E{{X}}+ \E{t\left(\phi(t-X)\right)+\frac{X}{\sqrt{2\pi}}} \\
 &\leq& t\E{{X}}+ \E{t\left(\phi(t)+\frac{2X}{t}\right)+\frac{X}{\sqrt{2\pi}}} \label{eq:phibnd2}\\
&\leq& \mathcal{E}t+ t\phi(t)+ \left(2+\frac{1}{\sqrt{2\pi}}\right)\mathcal{E},
% &=& \Econd{X^2+(t^2-X^2)\mathcal{Q}(t-X)}{X\leq t}\text{Pr}(X\leq t) + \Econd{t^2+(X^2-t^2)\mathcal{Q}(X-t)}{X > t}\text{Pr}(X > t) \\
%&\leq& \Econd{Xt+(t+X)(t-X)\mathcal{Q}(t-X)}{X\leq t}\text{Pr}(X\leq t) + \Econd{Xt+(X+t)(X-t)\mathcal{Q}(X-t)}{X > t}\text{Pr}(X > t) \nonumber\\ 
%\label{eq:EE2110}
%\\
%&\leq& \Econd{Xt+(t+X)\phi(t-X)}{X\leq t}\text{Pr}(X\leq t) + \Econd{Xt+(X+t)\phi(X-t)}{X > t}\text{Pr}(X > t) \\
%&=& t\E{X}+ \E{(t+X)\phi(t-X)} \\
%&\leq& \mathcal{E}t+ t\E{\phi(t-X)} + \mathcal{E} \\
%&\leq& \mathcal{E}t+ t\E{\phi(t)+\frac{2X}{t}} + \mathcal{E} \\
%&\leq& \mathcal{E}t+ t\phi(t) + 3\mathcal{E}.
\label{eq:EE222}
\end{IEEEeqnarray}
where~\eqref{eq:phibnd2} follows from Lemma~\ref{lemm2}.
\llg{
Further substituting~\eqref{eq:EE222} into~\eqref{eq:EE2112}, we have
\begin{IEEEeqnarray}{rCl}
\IEEEeqnarraymulticol{3}{l}{
   \E{-\int_{-\infty}^{\infty} W(y|X)\log R(y) \dd y }
} \nonumber\\*\quad
&\leq&  \frac{1}{2}\log 2\pi e + \frac{e^{-\frac{t^2}{2}}}{1-e^{-\frac{t^2}{2}}}+{\left(\frac{t}{2}+1\right)\phi(t)+\frac{2\mathcal{E}}{t}}+\left(2+\frac{1}{2\sqrt{2\pi}}\right)\mathcal{E}+ \frac{\mathcal{E}t}{2}. 
 \label{eq:finaexp}
\end{IEEEeqnarray}
Then, by~\eqref{eq:dualbnd} we have 
\begin{IEEEeqnarray}{rCl}
  \mathsf{C}_{\textnormal{G}} &\leq& 
 \frac{e^{-\frac{t^2}{2}}}{1-e^{-\frac{t^2}{2}}}+{\left(\frac{t}{2}+1\right)\phi(t)+\frac{2\mathcal{E}}{t}}+\left(2+\frac{1}{2\sqrt{2\pi}}\right)\mathcal{E}+ \frac{\mathcal{E}t}{2}.
 \label{eq:finaexp3} 
\end{IEEEeqnarray} }
\llg{Recalling} $t=a_{\textnormal{G}}\sqrt{\log\frac{1}{\mathcal{E}}}$, and \llg{substituting} it into~\eqref{eq:finaexp3}, we have
\begin{IEEEeqnarray}{rCl}
\frac{e^{-\frac{t^2}{2}}}{1-e^{-\frac{t^2}{2}}} &=& \frac{\mathcal{E}^{\frac{a_{\textnormal{G}}^2}{2}}}{1-\mathcal{E}^{\frac{a_{\textnormal{G}}^2}{2}}}~\dot{=}~ \mathcal{E}^{\frac{a_{\textnormal{G}}^2}{2}},\label{eq:firstt}\\
\left(\frac{t}{2}+1\right)\phi(t) &=& \left(\frac{a_{\textnormal{G}}}{2}\sqrt{\log\frac{1}{\mathcal{E}}}+1 \right) \frac{\mathcal{E}^{\frac{a_{\textnormal{G}}^2}{2}}}{\sqrt{2\pi}} ~\dot{=}~ \frac{a_{\textnormal{G}}}{2\sqrt{2\pi}}\mathcal{E}^{\frac{a_{\textnormal{G}}^2}{2}}\sqrt{\log\frac{1}{\mathcal{E}}}, \label{eq:secondt} \\
\frac{2\mathcal{E}}{t} &=& \frac{2}{a_{\textnormal{G}}}\frac{\mathcal{E}}{\sqrt{\log\frac{1}{\mathcal{E}}}}, \label{eq:thirdt} \\
 \frac{\mathcal{E} t}{2} &=& \frac{a_{\textnormal{G}}}{2}{\mathcal{E}\sqrt{\log\frac{1}{\mathcal{E}}}}. \label{eq:domin}
\end{IEEEeqnarray}
Comparing~\eqref{eq:firstt},~\eqref{eq:secondt}, and~\eqref{eq:thirdt} with~\eqref{eq:domin}, and recalling $a_{\textnormal{G}} > \sqrt{2}$, we can observe that the last term~\eqref{eq:domin} dominates for \llg{$\mathcal{E} \to 0^{+}$}. Hence, 
\begin{IEEEeqnarray}{rCl}
 \mathsf{C}_{\textnormal{G}}  ~\dot{\leq}~ \frac{a_{\textnormal{G}}}{2}\mathcal{E}\sqrt{\log\frac{1}{\mathcal{E}}}.
\end{IEEEeqnarray}
Since $a_{\textnormal{G}}>\sqrt{2}$ is chosen arbitrarily, 
\begin{IEEEeqnarray}{rCl}
 \mathsf{C}_{\textnormal{G}} ~\dot{\leq}~ \inf_{a_{\textnormal{G}}>\sqrt{2}} \frac{a_{\textnormal{G}}}{2}\mathcal{E}\sqrt{\log\frac{1}{\mathcal{E}}}  &~\dot{=}~&  \frac{1}{\sqrt{2}} \mathcal{E}\sqrt{\log\frac{1}{\mathcal{E}}}.
\end{IEEEeqnarray}
Eq.~\eqref{eq:11} is proved. 
\subsubsection{Proof of Eq.~\eqref{eq:12}}
\label{secteq12}
Eq.~\eqref{eq:12} was already proved in~\cite{lapidothmoserwigger09_7}, in which the proof involves quite complicated evaluations on several integral items. Here, we give a new simple proof \llg{based on the data processing inequality and Fano's inequality}. 

Consider a binary input $X_\text{B}$ with the distribution
\begin{IEEEeqnarray}{rCl}
p_{X_\text{B}}=\begin{cases}
 1-\frac{\mathcal{E}}{x_0} &\text{  if  } X_\text{B}=0, \\
 \frac{\mathcal{E}}{x_0} &\text{  if } X_\text{B}=x_0,
\end{cases}
\end{IEEEeqnarray}
where $x_0=a_{\textnormal{G}} \sqrt{\log \frac{1}{\mathcal{E}}}$ with $a_{\textnormal{G}}>\sqrt{2}$.

 Given $Y$, denote $\hat{X}_\text{B}$ as the estimate of $X_\text{B}$ by the maximum a posteriori probability (MAP) decision rule, i.e.,
\begin{IEEEeqnarray}{rCl}
\hat{X}_\text{B}=\argmax_{X}~\text{Pr}(X|Y).
\label{eq:maprule}
\end{IEEEeqnarray}   
Then the error probability $\text{P}_{\text{e}}$ by the MAP rule can be calculated as \llg{
\begin{IEEEeqnarray}{rCl}
\text{P}_{\text{e}} &=& 
\textnormal{Pr}(X_\text{B}=0)\textnormal{Pr}\left(Y>t\right)+\textnormal{Pr}(X_\text{B}=x_0)\textnormal{Pr}\left(Y \leq t\right)\\
&=& \left(1-\frac{\mathcal{E}}{x_0}\right)\Q\left(\frac{x_0}{2}+\frac{\log(\frac{x_0}{\mathcal{E}}-1)}{x_0}\right) +  \frac{\mathcal{E}}{x_0}\Q\left(\frac{x_0}{2}-\frac{\log(\frac{x_0}{\mathcal{E}}-1)}{x_0}\right),\label{eq:peexp2}
%\\
%%
%&\dot{=}& \left(1-\frac{\mathcal{E}}{x_0}\right) \frac{\phi\left(\frac{x_0}{2}+\frac{\log(\frac{x_0}{\mathcal{E}}-1)}{x_0}\right)}{\left(\frac{x_0}{2}+\frac{\log(\frac{x_0}{\mathcal{E}}-1)}{x_0}\right)} +  \frac{\mathcal{E}}{x_0}\frac{\phi\left(\frac{x_0}{2}-\frac{\log(\frac{x_0}{\mathcal{E}}-1)}{x_0}\right)}{\left(\frac{x_0}{2}-\frac{\log(\frac{x_0}{\mathcal{E}}-1)}{x_0}\right)}\\
%%&\dot{=}& \mathcal{Q}\left(\left(\frac{a}{2}+\frac{1}{a}\right)\sqrt{\log\frac{1}{\mathcal{E}}}\right) \\
%&\dot{=}& \frac{\mathcal{E}^{\frac{1}{2}\left(\frac{a}{2}+\frac{1}{a}\right)^2}}{\sqrt{2\pi}\left(\frac{a}{2}+\frac{1}{a}\right)\sqrt{\log\frac{1}{\mathcal{E}}}} + \frac{\mathcal{E}^{1+\frac{1}{2}\left(\frac{a}{2}-\frac{1}{a}\right)^2}}{\sqrt{2\pi}\left(\frac{a}{2}-\frac{1}{a}\right){\log\frac{1}{\mathcal{E}}}} \\
%&\dot{=}& \frac{\mathcal{E}^{\frac{1}{2}\left(\frac{a}{2}+\frac{1}{a}\right)^2}}{\sqrt{2\pi}\left(\frac{a}{2}+\frac{1}{a}\right)\sqrt{\log\frac{1}{\mathcal{E}}}}.
\end{IEEEeqnarray}
where $t=\frac{x_0}{2}+\frac{\log(\frac{x_0}{\mathcal{E}}-1)}{x_0}$, denotes the decision threshold of the likelihood ratio~\cite{Proakis2008}.

For the first term at the RHS of~\eqref{eq:peexp2}, \llg{ recalling $x_0=a_{\textnormal{G}} \sqrt{\log \frac{1}{\mathcal{E}}}$ and} 
\begin{IEEEeqnarray}{rCl}
\IEEEeqnarraymulticol{3}{l}{
 \bigg(\underbrace{1-\frac{\mathcal{E}}{x_0}}_{\leq 1}\bigg) \Q\left(\frac{x_0}{2}+\frac{\log\left(\frac{x_0}{\mathcal{E}}-1\right)}{x_0}\right) 
}\nonumber\\*\quad
 &\leq&\Q\left(\frac{x_0}{2}+\frac{\log\left(\frac{x_0}{\mathcal{E}}-1\right)}{x_0}\right)  \\
 &{=}& \Q\left(\frac{a_{\textnormal{G}} \sqrt{\log \frac{1}{\mathcal{E}}}}{2}+\frac{\log\left(\frac{a_{\textnormal{G}} \sqrt{\log \frac{1}{\mathcal{E}}}}{\mathcal{E}}-1\right)}{a_{\textnormal{G}} \sqrt{\log \frac{1}{\mathcal{E}}}}\right)  \\
 &{=}& \Q\left(\frac{a_{\textnormal{G}} \sqrt{\log \frac{1}{\mathcal{E}}}}{2}+\frac{\log\frac{1}{\mathcal{E}}\left({a_{\textnormal{G}} \sqrt{\log \frac{1}{\mathcal{E}}}}-\mathcal{E}\right)}{a_{\textnormal{G}} \sqrt{\log \frac{1}{\mathcal{E}}}}\right)  \\
 &{=}& \Q\left(\frac{a_{\textnormal{G}} \sqrt{\log \frac{1}{\mathcal{E}}}}{2}+\frac{\log\frac{1}{\mathcal{E}}+\log\left({a_{\textnormal{G}} \sqrt{\log \frac{1}{\mathcal{E}}}}-\mathcal{E}\right)}{a_{\textnormal{G}} \sqrt{\log \frac{1}{\mathcal{E}}}}\right)  \\
 &{=}& \Q\left( \left(\frac{a_{\textnormal{G}} }{2}+\frac{1}{a_{\textnormal{G}}}\right)\sqrt{\log \frac{1}{\mathcal{E}}}+\frac{\log\left({a_{\textnormal{G}} \sqrt{\log \frac{1}{\mathcal{E}}}}-\mathcal{E}\right)}{a_{\textnormal{G}} \sqrt{\log \frac{1}{\mathcal{E}}}}\right)  \\
 &\leq& \Q\left( \left(\frac{a_{\textnormal{G}} }{2}+\frac{1}{a_{\textnormal{G}}}\right)\sqrt{\log \frac{1}{\mathcal{E}}}\right) \label{eq:simplq} \\
 &\leq&
 \frac{\phi\left(\left(\frac{a_{\textnormal{G}} }{2}+\frac{1}{a_{\textnormal{G}}}\right)\sqrt{\log \frac{1}{\mathcal{E}}}\right)}{\left(\frac{a_{\textnormal{G}} }{2}+\frac{1}{a_{\textnormal{G}}}\right)\sqrt{\log \frac{1}{\mathcal{E}}}}
\label{eq:expan} \\
&~\dot{=}~& \frac{\mathcal{E}^{\frac{1}{2}\left(\frac{a_{\textnormal{G}}}{2}+\frac{1}{a_{\textnormal{G}}}\right)^2}}{\sqrt{2\pi}\left(\frac{a_{\textnormal{G}}}{2}+\frac{1}{a_{\textnormal{G}}}\right)\sqrt{\log\frac{1}{\mathcal{E}}}}, \label{eq:firstterm1}
\end{IEEEeqnarray}
where~\eqref{eq:simplq} follows from the fact that $\Q(x)$ decreases as $x$ increases, and~\eqref{eq:expan} follows from $\mathcal{Q}\left(x\right) \leq \frac{\phi\left(x\right)}{x},~x>0$. 

For the second term,
\begin{IEEEeqnarray}{rCl}
\IEEEeqnarraymulticol{3}{l}{
\frac{\mathcal{E}}{x_0}\Q\left(\frac{x_0}{2}-\frac{\log(\frac{x_0}{\mathcal{E}}-1)}{x_0}\right)
} \nonumber\\*\,
 &=&
\frac{\mathcal{E}}{a_{\textnormal{G}}\sqrt{\log\frac{1}{\mathcal{E}}}}\Q\left( \left(\frac{a_{\textnormal{G}} }{2}-\frac{1}{a_{\textnormal{G}}}\right)\sqrt{\log \frac{1}{\mathcal{E}}}-\frac{\log\left({a_{\textnormal{G}} \sqrt{\log \frac{1}{\mathcal{E}}}}-\mathcal{E}\right)}{a_{\textnormal{G}} \sqrt{\log \frac{1}{\mathcal{E}}}}\right)  \\
&\leq&\frac{\mathcal{E}}{a_{\textnormal{G}}\sqrt{\log\frac{1}{\mathcal{E}}}}\frac{\phi\left( \left(\frac{a_{\textnormal{G}} }{2}-\frac{1}{a_{\textnormal{G}}}\right)\sqrt{\log \frac{1}{\mathcal{E}}}-\frac{\log\left({a_{\textnormal{G}} \sqrt{\log \frac{1}{\mathcal{E}}}}-\mathcal{E}\right)}{a_{\textnormal{G}} \sqrt{\log \frac{1}{\mathcal{E}}}}\right)}{\left(\frac{a_{\textnormal{G}} }{2}-\frac{1}{a_{\textnormal{G}}}\right)\sqrt{\log \frac{1}{\mathcal{E}}}-\frac{\log\left({a_{\textnormal{G}} \sqrt{\log \frac{1}{\mathcal{E}}}}-\mathcal{E}\right)}{a_{\textnormal{G}} \sqrt{\log \frac{1}{\mathcal{E}}}}} \label{eq:ageq2} \\
&=&\frac{\mathcal{E}}{a_{\textnormal{G}}\sqrt{\log\frac{1}{\mathcal{E}}}}\frac{e^{-\frac{1}{2} \left(\frac{a_{\textnormal{G}} }{2}-\frac{1}{a_{\textnormal{G}}}\right)^2{\log \frac{1}{\mathcal{E}}} }e^{\frac{1}{a_{\textnormal{G}}}\left(\frac{a_{\textnormal{G}}}{2}-\frac{1}{a_{\textnormal{G}}}\right)\log\left({a_{\textnormal{G}} \sqrt{\log \frac{1}{\mathcal{E}}}}-\mathcal{E}\right)}\underbrace{e^{-\frac{1}{2} \left( \frac{\log\left({a_{\textnormal{G}} \sqrt{\log \frac{1}{\mathcal{E}}}}-\mathcal{E}\right)}{a_{\textnormal{G}} \sqrt{\log \frac{1}{\mathcal{E}}}} \right)^2 }}_{\dot{=}~1} }{\sqrt{2\pi}\left(  {\left(\frac{a_{\textnormal{G}} }{2}-\frac{1}{a_{\textnormal{G}}}\right)\sqrt{\log \frac{1}{\mathcal{E}}}-\frac{\log\left({a_{\textnormal{G}} \sqrt{\log \frac{1}{\mathcal{E}}}}-\mathcal{E}\right)}{a_{\textnormal{G}} \sqrt{\log \frac{1}{\mathcal{E}}}} }   \right)} \nonumber \\
\label{eq:definq}\\
%&=&\frac{\mathcal{E}}{a_{\textnormal{G}}\sqrt{\log\frac{1}{\mathcal{E}}}}\frac{\frac{1}{\sqrt{2\pi}}\left(e^{-\frac{1}{2}\left( \frac{a_{\textnormal{G}}}{2}-\frac{1}{a_{\textnormal{G}}}\right)^2\log\frac{1}{\mathcal{E}} } + e^{\frac{1}{a_{\textnormal{G}}}(\frac{a_{\textnormal{G}}}{2}-\frac{1}{a_{\textnormal{G}}})\log (a_{\textnormal{G}}\sqrt{\log \frac{1}{\mathcal{E}}}-\mathcal{E}) } + e^{-\frac{1}{2}\left(\frac{\log\left({a_{\textnormal{G}} \sqrt{\log \frac{1}{\mathcal{E}}}}-\mathcal{E}\right)}{a_{\textnormal{G}} \sqrt{\log \frac{1}{\mathcal{E}}}}\right)^2}   \right)   }{\left(\frac{a_{\textnormal{G}} }{2}-\frac{1}{a_{\textnormal{G}}}\right)\sqrt{\log \frac{1}{\mathcal{E}}}-\frac{\log\left({a_{\textnormal{G}} \sqrt{\log \frac{1}{\mathcal{E}}}}-\mathcal{E}\right)}{a_{\textnormal{G}} \sqrt{\log \frac{1}{\mathcal{E}}}}}  \\
%&~\dot{=}~&\frac{\mathcal{E}}{a_{\textnormal{G}}\sqrt{\log\frac{1}{\mathcal{E}}}}\frac{\phi\left( \left(\frac{a_{\textnormal{G}} }{2}-\frac{1}{a_{\textnormal{G}}}\right)\sqrt{\log \frac{1}{\mathcal{E}}}-\frac{\log\left({a_{\textnormal{G}} \sqrt{\log \frac{1}{\mathcal{E}}}}-\mathcal{E}\right)}{a_{\textnormal{G}} \sqrt{\log \frac{1}{\mathcal{E}}}}\right)}{\left(\frac{a_{\textnormal{G}} }{2}-\frac{1}{a_{\textnormal{G}}}\right)\sqrt{\log \frac{1}{\mathcal{E}}} }  \\
&~\dot{=}~& \frac{\mathcal{E}}{ a_{\textnormal{G}}\sqrt{\log\frac{1}{\mathcal{E}}}  } \frac{\mathcal{E}^{\frac{1}{2}\left(\frac{a_{\textnormal{G}}}{2}-\frac{1}{a_{\textnormal{G}}}\right)^2} \left( a_{\textnormal{G}}\sqrt{\log\frac{1}{\mathcal{E}}}-\mathcal{E} \right)^{\frac{1}{a_{\textnormal{G}}}\left(\frac{a_{\textnormal{G}}}{2}-\frac{1}{a_{\textnormal{G}}}\right)}  }{\sqrt{2\pi}\left(\frac{a_{\textnormal{G}}}{2}-\frac{1}{a_{\textnormal{G}}}\right){ \sqrt{\log\frac{1}{\mathcal{E}}}}} \label{eq:mm145} \\
&~\dot{=}~& \frac{\mathcal{E}}{ a_{\textnormal{G}}\sqrt{\log\frac{1}{\mathcal{E}}}  } \frac{\mathcal{E}^{\frac{1}{2}\left(\frac{a_{\textnormal{G}}}{2}-\frac{1}{a_{\textnormal{G}}}\right)^2} \left( a_{\textnormal{G}}\sqrt{\log\frac{1}{\mathcal{E}}} \right)^{\frac{1}{a_{\textnormal{G}}}\left(\frac{a_{\textnormal{G}}}{2}-\frac{1}{a_{\textnormal{G}}}\right)}  }{\sqrt{2\pi}\left(\frac{a_{\textnormal{G}}}{2}-\frac{1}{a_{\textnormal{G}}}\right){ \sqrt{\log\frac{1}{\mathcal{E}}}}} \\
&~\dot{=}~& \frac{\mathcal{E}^{\frac{1}{2}\left(\frac{a_{\textnormal{G}}}{2}+\frac{1}{a_{\textnormal{G}}}\right)^2}}{\sqrt{2\pi}\left(\frac{a_{\textnormal{G}}}{2}-\frac{1}{a_{\textnormal{G}}}\right){a_{\textnormal{G}}}^{\frac{1}{2}+\frac{1}{{a_{\textnormal{G}}}^2}} \left(\sqrt{\log\frac{1}{\mathcal{E}}}\right)^{\frac{3}{2} + \frac{1}{{a_{\textnormal{G}}}^2} } } ,
\label{eq:secondterm1}
\end{IEEEeqnarray}
where~\eqref{eq:ageq2} follows from $a_{\textnormal{G}}>\sqrt{2}$ and $\mathcal{Q}\left(x\right) \leq \frac{\phi\left(x\right)}{x},~x>0$, \eqref{eq:definq} from the definition~$\phi(x)=\frac{e^{-\frac{x^2}{2}}}{\sqrt{2\pi}}$, and~\eqref{eq:mm145} from the equation $e^{a \log x}=x^{a},~x>0$. 
\begin{remark}
It should be noted that the condition $a_{\textnormal{G}}>\sqrt{2}$ is necessary for the derivation of~\eqref{eq:ageq2}. To make~\eqref{eq:ageq2}, the parameter $ \left(\frac{a_{\textnormal{G}} }{2}-\frac{1}{a_{\textnormal{G}}}\right)\sqrt{\log \frac{1}{\mathcal{E}}}-\frac{\log\left({a_{\textnormal{G}} \sqrt{\log \frac{1}{\mathcal{E}}}}-\mathcal{E}\right)}{a_{\textnormal{G}} \sqrt{\log \frac{1}{\mathcal{E}}}} $ in $\mathcal{Q}(\cdot)$ needs to be positive when $\mathcal{E}$ is small enough. This can be satisfied by letting $\left(\frac{a_{\textnormal{G}} }{2}-\frac{1}{a_{\textnormal{G}}}\right) >0$, which is equivalent to  $a_{\textnormal{G}}>\sqrt{2}$.
\end{remark} }
\llg{Substituting~\eqref{eq:firstterm1} and~\eqref{eq:secondterm1} into~\eqref{eq:peexp2} yields}
\begin{IEEEeqnarray}{rCl}
\text{P}_{\text{e}} 
&~\dot{\leq}~& \frac{\mathcal{E}^{\frac{1}{2}\left(\frac{a_{\textnormal{G}}}{2}+\frac{1}{a_{\textnormal{G}}}\right)^2}}{\sqrt{2\pi}\left(\frac{a_{\textnormal{G}}}{2}+\frac{1}{a_{\textnormal{G}}}\right)\sqrt{\log\frac{1}{\mathcal{E}}}} +
 \frac{\mathcal{E}^{\frac{1}{2}\left(\frac{a_{\textnormal{G}}}{2}+\frac{1}{a_{\textnormal{G}}}\right)^2}}{\sqrt{2\pi}\left(\frac{a_{\textnormal{G}}}{2}-\frac{1}{a_{\textnormal{G}}}\right){a_{\textnormal{G}}}^{\frac{1}{2}+\frac{1}{{a_{\textnormal{G}}}^2}} \left(\sqrt{\log\frac{1}{\mathcal{E}}}\right)^{\frac{3}{2} + \frac{1}{{a_{\textnormal{G}}}^2} } } \\
&~\dot{=}~& \frac{\mathcal{E}^{\frac{1}{2}\left(\frac{a_{\textnormal{G}}}{2}+\frac{1}{a_{\textnormal{G}}}\right)^2}}{\sqrt{2\pi}\left(\frac{a_{\textnormal{G}}}{2}+\frac{1}{a_{\textnormal{G}}}\right)\sqrt{\log\frac{1}{\mathcal{E}}}}, \label{eq:finalpe2}
\end{IEEEeqnarray}
where~\eqref{eq:finalpe2} follows from the fact that the first term dominates for $\mathcal{E} \to 0^{+}$.

Since $X_\text{B}-Y-\hat{X}_\text{B}$ forms a Markov chain, by the data processing inequality, $\mathsf{I}(X_\text{B};Y)$ can be lower-bounded by
\begin{IEEEeqnarray}{rCl}
\mathsf{I}(X_\text{B};Y) 
 &\geq& \mathsf{I}(X_\text{B};\hat{X}_\text{B}) \label{eq:dpi} \\
&=& \mathsf{H}(X_\text{B})-\mathsf{H}(X_\text{B}|\hat{X}_\text{B}) \\
&\geq& \mathsf{H}(X_\text{B})-\mathsf{H}_{\text{b}}(\text{P}_{\text{e}}), \label{eq:fi}
\end{IEEEeqnarray}
where~$\mathsf{H}_{\text{b}}(\text{P}_{\text{e}}) \overset{\text{def}}{=}  -\text{P}_{\text{e}}\log \text{P}_{\text{e}} -(1-\text{P}_{\text{e}})\log (1-\text{P}_{\text{e}})$, and \eqref{eq:fi} follows by the Fano's inequality. 

\llg{Recalling again $x_0=a_{\textnormal{G}} \sqrt{\log \frac{1}{\mathcal{E}}}$, we obtain} 
\begin{IEEEeqnarray}{rCl}
\mathsf{H}(X_\text{B}) &=& -\frac{\mathcal{E}}{x_0}\log\frac{\mathcal{E}}{x_0}-\left(1-\frac{\mathcal{E}}{x_0}\right)\log\left(1-\frac{\mathcal{E}}{x_0}\right) \\
&\dot{=}& -\frac{\mathcal{E}}{x_0}\log\frac{\mathcal{E}}{x_0} \\
&\dot{=}& \frac{\mathcal{E}}{a_{\textnormal{G}} \sqrt{\log \frac{1}{\mathcal{E}}}} \log \frac{a_{\textnormal{G}} \sqrt{\log \frac{1}{\mathcal{E}}}}{\mathcal{E}} \\
&\dot{=}& \frac{\mathcal{E}}{a_{\textnormal{G}} \sqrt{\log \frac{1}{\mathcal{E}}}}\left(\log\frac{1}{\mathcal{E}}+\log a_{\textnormal{G}}+\frac{1}{2}\log\log\frac{1}{\mathcal{E}} \right) \label{eq:wq468} \\
 &\dot{=}& \frac{1}{a_{\textnormal{G}}} \mathcal{E}\sqrt{\log\frac{1}{\mathcal{E}}}.
\label{eq:EEEE2003}
\end{IEEEeqnarray}
where~\eqref{eq:wq468} follows from the fact that the first term dominates for $\mathcal{E} \to 0^{+}$.

We bound the term $\mathsf{H}_{\text{b}}(\text{P}_{\text{e}})$ by
\begin{IEEEeqnarray}{rCl}
\mathsf{H}_{\text{b}}(\text{P}_{\text{e}}) &=&  -\text{P}_{\text{e}}\log \text{P}_{\text{e}} -(1-\text{P}_{\text{e}})\log (1-\text{P}_{\text{e}}) \\
%&\dot{=}& -\text{P}_{\text{e}}\log \text{P}_{\text{e}} + \text{P}_{\text{e}} -{\text{P}^2_e} \\
&\dot{=}& -\text{P}_{\text{e}}\log \text{P}_{\text{e}} \\
 &\dot{\leq}& \frac{\mathcal{E}^{\frac{1}{2}\left(\frac{a_{\textnormal{G}}}{2}+\frac{1}{a_{\textnormal{G}}}\right)^2}}{\sqrt{2\pi}\left(\frac{a_{\textnormal{G}}}{2}+\frac{1}{a_{\textnormal{G}}}\right)\sqrt{\log\frac{1}{\mathcal{E}}}} \left(\log\frac{1}{\mathcal{E}}+\log \sqrt{2\pi}\left(\frac{a_{\textnormal{G}}}{2}+\frac{1}{a_{\textnormal{G}}}\right)+\frac{1}{2}\log\log\frac{1}{\mathcal{E}} \right) \label{eq:sumterm2} 
\IEEEeqnarraynumspace\\
&\dot{=}&  \frac{1}{\sqrt{2\pi}\left(\frac{a_{\textnormal{G}}}{2}+\frac{1}{a_{\textnormal{G}}}\right)\sqrt{\log\frac{1}{\mathcal{E}}}}\mathcal{E}^{\frac{1}{2}\left(\frac{a_{\textnormal{G}}}{2}+\frac{1}{a_{\textnormal{G}}}\right)^2} \log\frac{1}{\mathcal{E}} \label{eq:supterm3}\\
&\dot{=}&  \frac{1}{\sqrt{2\pi}\left(\frac{a_{\textnormal{G}}}{2}+\frac{1}{a_{\textnormal{G}}}\right)}\mathcal{E}^{\frac{1}{2}\left(\frac{a_{\textnormal{G}}}{2}+\frac{1}{a_{\textnormal{G}}}\right)^2} \sqrt{ \log\frac{1}{\mathcal{E}}},
\label{eq:EEEE2013}
\end{IEEEeqnarray}
\llg{where~\eqref{eq:sumterm2} follows by~\eqref{eq:finalpe2}, and~\eqref{eq:supterm3} by the fact that the first term dominates for $\mathcal{E} \to 0^{+}$. 

Substituting~\eqref{eq:EEEE2003} and~\eqref{eq:EEEE2013} into~\eqref{eq:fi}, we obtain}
\begin{IEEEeqnarray}{rCl}
\mathsf{I}(X_\text{B};Y) 
&~\dot{\geq}~& \frac{1}{a_{\textnormal{G}}} \mathcal{E}\sqrt{\log\frac{1}{\mathcal{E}}} -   \frac{1}{\sqrt{2\pi}\left(\frac{a_{\textnormal{G}}}{2}+\frac{1}{a_{\textnormal{G}}}\right)}\mathcal{E}^{\frac{1}{2}\left(\frac{a_{\textnormal{G}}}{2}+\frac{1}{a_{\textnormal{G}}}\right)^2} \sqrt{ \log\frac{1}{\mathcal{E}}}\\
&~\dot{=}~& \frac{1}{a_{\textnormal{G}}} \mathcal{E}\sqrt{\log\frac{1}{\mathcal{E}}}, \label{eq:finalupp2}
\end{IEEEeqnarray}
where~\eqref{eq:finalupp2} follows from the fact that $\frac{1}{2}\left(\frac{a_{\textnormal{G}}}{2}+\frac{1}{a_{\textnormal{G}}}\right)^2 > 1$ when $a_{\textnormal{G}}>\sqrt{2}$, and hence the first term dominates for $\mathcal{E} \to 0^{+}$.

Then, the capacity can be lower-bounded by
\begin{IEEEeqnarray}{rCl}
\mathsf{C}_{\textnormal{G}} \geq \mathsf{I}(X_\text{B};Y)
~\dot{\geq}~  \frac{1}{a_{\textnormal{G}}} \mathcal{E}\sqrt{\log\frac{1}{\mathcal{E}}}.
\end{IEEEeqnarray}
Since $a_{\textnormal{G}}>\sqrt{2}$ is chosen arbitrarily, 
\begin{IEEEeqnarray}{rCl}
\mathsf{C}_{\textnormal{G}}
~\dot{\geq}~ \sup_{a_{\textnormal{G}}>\sqrt{2}} \frac{1}{a_{\textnormal{G}}} \mathcal{E}\sqrt{\log\frac{1}{\mathcal{E}}} 
&~\dot{=}~& \frac{1}{\sqrt{2}}\mathcal{E}\sqrt{\log\frac{1}{\mathcal{E}}}.
\end{IEEEeqnarray}
Eq.~\eqref{eq:12} is proved. 

%\end{IEEEproof}
%In this note we are interested in characterizing the constant
%\begin{IEEEeqnarray}{rCl}
%\alpha = \lim_{\mathcal{E} \to 0^{+}} \frac{\mathsf{C}(\mathcal{E})}{\mathcal{E}{\log \log \frac{1}{\mathcal{E}} }}.
%\end{IEEEeqnarray} 
%\begin{lemma}
%Given $u,v>0$, 
%\begin{IEEEeqnarray}{rCl}
%\mathcal{Q}(-v)-\frac{u}{2v}  \leq \mathcal{Q}(u-v) \leq \mathcal{Q}(u)+ \frac{v}{2u}.
%\end{IEEEeqnarray}
%\end{lemma}
%\begin{IEEEproof}
%Let
%\end{IEEEproof}
\subsection{Poisson Optical Intensity Channel}
\begin{theorem}
\label{thm2}
The capacity of channel~\eqref{eq:singlechannelmodel2} satisfies
\begin{IEEEeqnarray}{rCl}
\lim_{\mathcal{E} \to 0^{+}} 
\frac{\mathsf{C}_{\textnormal{P}}(\mathcal{E})}{ {\mathcal{E}{\log \log \frac{1}{\mathcal{E}} }} }  = 1.
\end{IEEEeqnarray} 
\end{theorem}
%\begin{IEEEproof}
\llg{We also prove Theorem~\ref{thm2} in two steps. It is equivalent to prove 
\begin{IEEEeqnarray}{rCl}
\limsup_{\mathcal{E} \to 0^{+}} 
\frac{\mathsf{C}_{\textnormal{P}}(\mathcal{E})}{ {\mathcal{E}{\log \log \frac{1}{\mathcal{E}} }} } \leq 1 ,
\label{eq:111}
\end{IEEEeqnarray}
and 
\begin{IEEEeqnarray}{rCl}
%\liminf_{\mathcal{E} \to 0^{+}^+}
 \liminf_{\mathcal{E} \to 0^{+}} 
\frac{\mathsf{C}_{\textnormal{P}}(\mathcal{E})}{ {\mathcal{E}{\log \log \frac{1}{\mathcal{E}} }} } \geq 1.
\label{eq:122}
\end{IEEEeqnarray}
We will prove Eq.~\eqref{eq:111} in Section~\ref{secteq111}, and prove Eq.~\eqref{eq:122} in Section~\ref{secteq122}.}
\subsubsection{Proof of Eq.~\eqref{eq:111}}
\label{secteq111}
\llg{We again use the duality-based upper bound on capacity}~\eqref{eq:dualbndorg}. The auxiliary distribution $R(\cdot)$ here is chosen as\footnote{We can verify that $R(\cdot)$ is a distribution by showing $\sum_{y=0}^{\infty}R(y) = \frac{1-\beta}{T_{\eta}}\sum_{y=0}^{\eta-1} \text{Poi}_{\lambda}(y) +\sum_{y=\eta}^{\infty}\beta (1-p)p^{y-\eta} = \frac{1-\beta}{\sum_{y=0}^{\eta-1} \text{Poi}_{\lambda}(y)}\sum_{y=0}^{\eta-1} \text{Poi}_{\lambda}(y) +\beta(1-p)\frac{1}{1-p}=1$. }
\begin{IEEEeqnarray}{rCl}
R(y) = 
\begin{cases}
\frac{1-\beta}{T_{\eta}}\text{Poi}_{\lambda}(y), \quad &y \in \{0,1,\ldots,\eta-1 \}, \\
\beta (1-p)p^{y-\eta}, \quad &y \in \{\eta,\eta+1,\ldots\},
\end{cases}
\label{eq:ry}
\end{IEEEeqnarray}
where ~$p \in (0,1)$ is a free parameter, \llg{$\eta$ denotes the largest integer that is less than or equal to the unique solution to $(\eta-\lambda)\log\frac{\eta}{\lambda}=a_{\textnormal{P}} \log\frac{1}{\mathcal{E}}$ with $a_{\textnormal{P}}>1$,~$\beta =e^{-(\eta-\lambda)\log\frac{\eta}{\lambda}}$}, and~$T_{\eta} = \sum_{y=0}^{\eta-1}\text{Poi}_{\lambda}(y)$. 
%It is direct to see $\eta\dot{=}\frac{a\log\frac{1}{\mathcal{E}}}{\log\log\frac{1}{\mathcal{E}}}$.

Substituting~\eqref{eq:ry} into the expectation term at the RHS of~\eqref{eq:E101} \llg{yields}
\begin{IEEEeqnarray}{rCl}
 \mathsf{C}_{\textnormal{P}}(\mathcal{E}) &\leq& \sup_{p_X} \mathsf{E} \Bigg[\underbrace{\sum_{y=0}^{\eta-1} \text{Poi}_{\lambda+X}(y)\log \frac{\text{Poi}_{\lambda+X}(y)}{R(y)}}_{c_1(X)}+\underbrace{\sum_{y=\eta}^{\infty} \text{Poi}_{\lambda+X}(y)\log \frac{\text{Poi}_{\lambda+X}(y)}{R(y)}}_{c_2(X)} \Bigg]. \IEEEeqnarraynumspace
\label{eq:mas123123}
\end{IEEEeqnarray}
In the following, we respectively upper-bound $c_1(X)$ and $c_2(X)$. For $c_1(X)$, \llg{
\begin{IEEEeqnarray}{rCl}
    c_1(X)&=&\sum_{y=0}^{\eta-1} \text{Poi}_{\lambda+X}(y)\log \frac{\text{Poi}_{\lambda+X}(y)}{R(y)} \\
  & = &  \sum_{y=0}^{\eta-1}
  \text{Poi}_{\lambda+X}(y)
  \left(\log{\frac{T_{\eta}}{1-\beta}} - X+y\log\left(1+\frac{X}{\lambda}\right)\right) \\
  & = &  \sum_{y=0}^{\eta-1}
  \text{Poi}_{\lambda+X}(y)
  \bigg(-\log{(1-\beta)}+ \underbrace{\log T_{\eta}- X}_{\leq 0}+y\log\left(1+\frac{X}{\lambda}\right)\bigg) \\
  & \leq &  
  -\log{(1-\beta)}\underbrace{\sum_{y=0}^{\eta-1}
  \text{Poi}_{\lambda+X}(y)}_{\leq 1} + \log\left(1+\frac{X}{\lambda}\right) \sum_{y=1}^{\eta-1}y
  \text{Poi}_{\lambda+X}(y) \\
%&\leq& -\log (1-\beta) + \log\left(1+\frac{X}{\lambda}\right)\sum_{y=0}^{\eta-1}
%  y\text{Poi}_{\lambda+X}(y)
%%&\dot{=}& \frac{1}{2}\log 2\pi e + \mathcal{E}^{\frac{a^2}{2}}+  \frac{1}{2}\E{X^2\mathcal{Q}(X-t)}.
%%&\leq&  {-\log \frac{1-\beta}{\sqrt{2\pi e}\mathcal{Q}(-t)}}
%%  + \frac{\mathcal{E}(1+t)}{2},
%\label{eq:add1}
%\\
&\leq& -\log (1-\beta) + \log\left(1+\frac{X}{\lambda}\right) \sum_{y=1}^{\eta-1} \underbrace{y\text{Poi}_{\lambda+X}(y)}_{=(\lambda+X)\textnormal{Poi}_{\lambda+X}(y-1)}
%&\dot{=}& \frac{1}{2}\log 2\pi e + \mathcal{E}^{\frac{a^2}{2}}+  \frac{1}{2}\E{X^2\mathcal{Q}(X-t)}.
%&\leq&  {-\log \frac{1-\beta}{\sqrt{2\pi e}\mathcal{Q}(-t)}}
%  + \frac{\mathcal{E}(1+t)}{2},
\\
&=& -\log (1-\beta) + (\lambda+X)\log\left(1+\frac{X}{\lambda}\right)\underbrace{\sum_{y=0}^{\eta-2}
  \text{Poi}_{\lambda+X}(y)}_{\leq \sum_{y=0}^{\eta-1}
  \text{Poi}_{\lambda+X}(y)}
%&\dot{=}& \frac{1}{2}\log 2\pi e + \mathcal{E}^{\frac{a^2}{2}}+  \frac{1}{2}\E{X^2\mathcal{Q}(X-t)}.
%&\leq&  {-\log \frac{1-\beta}{\sqrt{2\pi e}\mathcal{Q}(-t)}}
%  + \frac{\mathcal{E}(1+t)}{2},
\label{eq:add11}
\\
&\leq& -\log (1-\beta) + \lambda\log\left(1+\frac{X}{\lambda}\right)\underbrace{\sum_{y=0}^{\eta-1}
  \text{Poi}_{\lambda+X}(y)}_{\leq 1} +X\log\left(1+\frac{X}{\lambda}\right)\sum_{y=0}^{\eta-1}
  \text{Poi}_{\lambda+X}(y) 
\nonumber \\
\\
%&\dot{=}& \frac{1}{2}\log 2\pi e + \mathcal{E}^{\frac{a^2}{2}}+  \frac{1}{2}\E{X^2\mathcal{Q}(X-t)}.
%&\leq&  {-\log \frac{1-\beta}{\sqrt{2\pi e}\mathcal{Q}(-t)}}
%  + \frac{\mathcal{E}(1+t)}{2},
\label{eq:add12}
%\\
%&\leq& -\log (1-\beta) + \lambda\log\left(1+\frac{\mathcal{E}}{\lambda}\right) + X\log\left(1+\frac{X}{\lambda}\right)\sum_{y=0}^{\eta-1}
%  \text{Poi}_{\lambda+X}(y).
%%&\dot{=}& \frac{1}{2}\log 2\pi e + \mathcal{E}^{\frac{a^2}{2}}+  \frac{1}{2}\E{X^2\mathcal{Q}(X-t)}.
%%&\leq&  {-\log \frac{1-\beta}{\sqrt{2\pi e}\mathcal{Q}(-t)}}
%%  + \frac{\mathcal{E}(1+t)}{2},
%\label{eq:add12}
&\leq& -\log (1-\beta) + \lambda\log\left(1+\frac{X}{\lambda}\right) +X\log\left(1+\frac{X}{\lambda}\right)\sum_{y=0}^{\eta-1}
  \text{Poi}_{\lambda+X}(y) .
\end{IEEEeqnarray}
%where $\phi(x) \eqdef \frac{1}{\sqrt{2\pi}} \ope^{-\frac{x^2}{2}}$, and eq.~\eqref{eq:EE207} holds by  $x\mathcal{Q}(x-t)=(x-t)\mathcal{Q}(x-t)+t\mathcal{Q}(x-t)\leq \phi(x-t)+t\leq 1+t$.
For $c_2(X)$,
\begin{IEEEeqnarray}{rCl}
  \IEEEeqnarraymulticol{3}{l}{%
    c_2(X)
  }\nonumber\\%
&=&\sum_{y=\eta}^{\infty} \text{Poi}_{\lambda+X}(y)\log \frac{\text{Poi}_{\lambda+X}(y)}{R(y)}\\
  & = &  \sum_{y=\eta}^{\infty}
  \text{Poi}_{\lambda+X}(y)
  \Bigg( -\log(\beta(1-p)) -(\underbrace{\lambda+X}_{\geq \lambda})-\eta \log \frac{1}{p} -\log {y!} + y\log \frac{\lambda+X}{p} \Bigg) \\
  & \leq &  \sum_{y=\eta}^{\infty}
  \text{Poi}_{\lambda+X}(y)
  \Bigg( -\log(\beta(1-p)) -\lambda-\eta \log \frac{1}{p} - \underbrace{\log y!}_{  \geq \log\left( \sqrt{2\pi y} (\frac{y}{e})^y\right) } + y\log \frac{\lambda+X}{p} \Bigg) 
\IEEEeqnarraynumspace\\
%&\leq& \sum_{y=\eta}^{\infty}
%  \text{Poi}_{\lambda+X}(y)
%  \left( -\log(\beta(1-p))-\eta \log \frac{1}{p}  -\log y! \right)+ \log\frac{\lambda+X}{p} \sum_{y=\eta}^{\infty}
%  y\text{Poi}_{\lambda+X}(y) \nonumber\\
&\leq& \sum_{y=\eta}^{\infty}
  \text{Poi}_{\lambda+X}(y)
  \Bigg( -\log(\beta(1-p))-\lambda-\eta \log \frac{1}{p} -\frac{1}{2}\underbrace{\log(2\pi y)}_{\geq \log(2\pi \eta)}-y\log y + y \nonumber\\
&&\hspace{2.6cm}+ y\log \frac{\lambda+X}{p}\Bigg) 
\label{eq:add2}
%\\
%&=& \sum_{y=\eta}^{\infty}
%  \text{Poi}_{\lambda+X}(y)
%  \left( -\log(\beta(1-p))-(\lambda+X)-\eta \log \frac{1}{p}  -y\log y \right) \nonumber\\
%&&\hspace{5.5cm}+ (\lambda+X)\left(1+\log\frac{\lambda+X}{p}\right) \sum_{y=\eta-1}^{\infty}
%  \text{Poi}_{\lambda+X}(y) \nonumber\\
%\label{eq:add3}
\\
&\leq& -\log\beta\sum_{y=\eta}^{\infty}
  \text{Poi}_{\lambda+X}(y) + {\left(-\log(1-p)-\lambda-\eta\log\frac{1}{p}-\frac{1}{2}\log (2\pi \eta)\right)} \sum_{y=\eta}^{\infty}
  \text{Poi}_{\lambda+X}(y)    \nonumber\\
&&+\left(1+\log\frac{1}{p}\right)\sum_{y=\eta}^{\infty} {y
  \text{Poi}_{\lambda+X}(y)}+ \underbrace{\sum_{y=\eta}^{\infty}
  (\log(\lambda+X)-\log y)y\text{Poi}_{\lambda+X}(y)}_{\leq 0}
%   \nonumber\\
\label{eq:add4123}
\\
&\leq& -\log\beta\sum_{y=\eta}^{\infty}
  \text{Poi}_{\lambda+X}(y) + {\left(-\log(1-p)-\lambda-\eta\log\frac{1}{p}-\frac{1}{2}\log (2\pi \eta)\right)} \sum_{y=\eta}^{\infty}
  \text{Poi}_{\lambda+X}(y)    \nonumber\\
&&+\left(1+\log\frac{1}{p}\right)\sum_{y=\eta}^{\infty} {y
  \text{Poi}_{\lambda+X}(y)},
%   \nonumber\\
\label{eq:add49}
\end{IEEEeqnarray}
where~\eqref{eq:add2} follows from Stirling's bound: $y!   \geq  \sqrt{2\pi y} (\frac{y}{e})^y $ . Eq.~\eqref{eq:add49} can be shown as follows: when $X< \eta-\lambda$, the last term at the RHS of~\eqref{eq:add4123} is negative because $\log(\lambda+X)-\log y <\log \eta -\log y \leq 0$, and when $X \geq \eta-\lambda$,
\begin{IEEEeqnarray}{rCl}
\IEEEeqnarraymulticol{3}{l}{
\sum_{y=\eta}^{\infty}
  (\log(\lambda+X)-\log y)y\text{Poi}_{\lambda+X}(y)   
}\nonumber\\*\quad
&=& \sum_{y=\eta}^{\infty}
  \log{\left(1+ \frac{\lambda+X-y}{y} \right) } \cdot y\text{Poi}_{\lambda+X}(y) \\
&\leq& \sum_{y=\eta}^{\infty}
  \frac{\lambda+X-y}{y}\cdot y\text{Poi}_{\lambda+X}(y)  \label{eq:m123d}\\
&=&  \sum_{y=\eta}^{\infty}
  ({\lambda+X})\text{Poi}_{\lambda+X}(y) -\sum_{y=\eta}^{\infty}
  \underbrace{{y}\text{Poi}_{\lambda+X}(y)}_{(\lambda+X)\textnormal{Poi}_{\lambda+X}(y-1)} 
\IEEEeqnarraynumspace
\\
&=& -(\lambda+X)\text{Poi}_{\lambda+X}(\eta-1) \\*
&\leq&0,
\end{IEEEeqnarray}
where~\eqref{eq:m123d} follows from $\log(1+x) \leq x,\, x>0$. Hence, the last term at the RHS of~\eqref{eq:add4123} is always nonpositive.

Continuing from~\eqref{eq:add49}, we have
\begin{IEEEeqnarray}{rCl}
\IEEEeqnarraymulticol{3}{l}{c_2(X)
}  \nonumber\\*\,
&\leq& -\log\beta\sum_{y=\eta}^{\infty}
  \text{Poi}_{\lambda+X}(y) + {\left(-\log(1-p)-\lambda-\eta\log\frac{1}{p}-\frac{1}{2}\log (2\pi \eta)\right)} \sum_{y=\eta}^{\infty}
  \text{Poi}_{\lambda+X}(y)    \nonumber\\
&&+\left(1+\log\frac{1}{p}\right)\sum_{y=\eta}^{\infty} \underbrace{y
  \text{Poi}_{\lambda+X}(y)}_{=(\lambda+X)\textnormal{Poi}_{\lambda+X}(y-1)}
%   \nonumber\\
%\label{eq:add49}
\\
&=& -\log\beta\sum_{y=\eta}^{\infty}
  \text{Poi}_{\lambda+X}(y) + {\left(-\log(1-p)-\lambda-\eta\log\frac{1}{p}-\frac{1}{2}\log (2\pi \eta)\right)} \sum_{y=\eta}^{\infty}
  \text{Poi}_{\lambda+X}(y)    \nonumber\\
&&+\left(1+\log\frac{1}{p}\right)(\lambda+X)\sum_{y=\eta-1}^{\infty}
  \text{Poi}_{\lambda+X}(y)
%   \nonumber\\
\label{eq:add4}
\\
&=& -\log\beta\sum_{y=\eta}^{\infty}
  \text{Poi}_{\lambda+X}(y) + {\left(-\log(1-p)-\lambda-\eta\log\frac{1}{p}-\frac{1}{2}\log (2\pi \eta)\right)} \sum_{y=\eta}^{\infty}
  \text{Poi}_{\lambda+X}(y)    \nonumber\\
&&+\lambda\left(1+\log\frac{1}{p}\right)\sum_{y=\eta-1}^{\infty}
  \text{Poi}_{\lambda+X}(y)+\left(1+\log\frac{1}{p}\right)X\underbrace{\sum_{y=\eta-1}^{\infty}
  \text{Poi}_{\lambda+X}(y)}_{\leq 1}
%   \nonumber\\
\label{eq:add499}
\\
&\leq& -\log\beta\sum_{y=\eta}^{\infty}
  \text{Poi}_{\lambda+X}(y) + {\left(-\log(1-p)-\lambda-\eta\log\frac{1}{p}-\frac{1}{2}\log (2\pi \eta)\right)} \sum_{y=\eta}^{\infty}
  \text{Poi}_{\lambda+X}(y)    \nonumber\\
&&+\lambda\left(1+\log\frac{1}{p}\right) \underbrace{\sum_{y=\eta-1}^{\infty}
  \text{Poi}_{\lambda+X}(y)}_{\textnormal{Poi}_{\lambda+X}(\eta-1)+\sum_{y=\eta}^{\infty}\textnormal{Poi}_{\lambda+X}(y)}+\left(1+\log\frac{1}{p}\right)X
%   \nonumber\\
\label{eq:add4999}
\\
&=& -\log\beta\sum_{y=\eta}^{\infty}
  \text{Poi}_{\lambda+X}(y) + {\left(-\log(1-p)-\lambda-\eta\log\frac{1}{p}-\frac{1}{2}\log (2\pi \eta)\right)} \sum_{y=\eta}^{\infty}
  \text{Poi}_{\lambda+X}(y)    \nonumber\\
&&+\lambda\left(1+\log\frac{1}{p}\right)
  \text{Poi}_{\lambda+X}(\eta-1)+\lambda\left(1+\log\frac{1}{p}\right)\sum_{y=\eta}^{\infty}
  \text{Poi}_{\lambda+X}(y)+\left(1+\log\frac{1}{p}\right)X
  \nonumber\\
\label{eq:add49999}
\\
&=& -\log\beta\sum_{y=\eta}^{\infty}
  \text{Poi}_{\lambda+X}(y) + \lambda\left(1+\log\frac{1}{p}\right)
  \text{Poi}_{\lambda+X}(\eta-1)+ \left(1+\log\frac{1}{p}\right)X   \nonumber\\
&&+\underbrace{\left(-\log(1-p)-\lambda-\eta\log\frac{1}{p}-\frac{1}{2}\log (2\pi \eta)+\lambda\left(1+\log\frac{1}{p}\right)\right)}_{\leq 0} \sum_{y=\eta}^{\infty}
  \text{Poi}_{\lambda+X}(y) 
\label{eq:md12313}
\\
&\leq& -\log\beta\sum_{y=\eta}^{\infty}
  \text{Poi}_{\lambda+X}(y) + \lambda\left(1+\log\frac{1}{p}\right)
  \text{Poi}_{\lambda+X}(\eta-1)+ \left(1+\log\frac{1}{p}\right)X
\label{eq:md12311}
\\
&\leq& -\log\beta\sum_{y=\eta}^{\infty}
  \text{Poi}_{\lambda+X}(y) + \lambda\left(1+\log\frac{1}{p}\right)\left(
  \text{Poi}_{\lambda}(\eta-1)+\frac{\text{Poi}_{\eta-2}(\eta-1)}{\eta-\lambda-2}X\right)
\nonumber\\
&&  +\left(1+\log\frac{1}{p}\right)X
\label{eq:addm1245}
\\
&=& -\log\beta\sum_{y=\eta}^{\infty}
  \text{Poi}_{\lambda+X}(y) + \lambda\left(1+\log\frac{1}{p}\right)
  \text{Poi}_{\lambda}(\eta-1)
\nonumber\\
&&  + \left(1+\log\frac{1}{p}\right)\left(1+\frac{\text{Poi}_{\eta-2}(\eta-1)}{\eta-\lambda-2}\lambda\right)X.
\label{eq:mmm149}
%  & \leq &
%\E{ \log {\frac{\mu}{\beta}}\cdot\Q(t-x) +\> \frac{x}{\mu}+\frac{\phi(t)+\frac{x}{\sqrt{2\pi}}}{{\mu}}  } \\
%& \leq &\E{ \log {\frac{\mu}{\beta}}\cdot\Q(t-x)} + \frac{2\mathcal{E}}{\mu}+\frac{\phi(t)}{\mu}.
%  & = &
% \log {\frac{\mu}{\beta}}\cdot (1-\Q(x-t)) +\> \frac{x}{\mu}+\frac{1}{\sqrt{2\pi}\mu}.
\end{IEEEeqnarray}
Here, Eq.~\eqref{eq:md12311} follows from the fact that $\lambda$ and $p$ are fixed constants, while the terms containing $\eta$ are all negative, and $\eta$ can be large enough by letting $\mathcal{E}$ small enough, to make the last term at the RHS of~\eqref{eq:md12313} being negative. Eq.~\eqref{eq:addm1245} can be derived by the following argument:
\begin{IEEEeqnarray}{rCl}
\text{Poi}_{\lambda+X}(\eta-1) &=&  \text{Poi}_{\lambda}(\eta-1) + \frac{\text{Poi}_{\lambda+X}(\eta-1)-\text{Poi}_{\lambda}(\eta-1)}{X}X \\
&\leq&  \text{Poi}_{\lambda}(\eta-1) + \frac{\text{Poi}_{\lambda+X}(\eta-1)}{X}X \\
&\leq&  \text{Poi}_{\lambda}(\eta-1) + \sup_{X>0} \left\{ \frac{\text{Poi}_{\lambda+X}(\eta-1)}{X} \right\} X \label{eq:supachi1}\\
&=&  \text{Poi}_{\lambda}(\eta-1) +  \frac{\text{Poi}_{\eta-2}(\eta-1)}{\eta-\lambda-2}  X,\label{eq:supachi2}
\end{IEEEeqnarray}
where~\eqref{eq:supachi2} follows by the supremum in~\eqref{eq:supachi1} being achieved at point $\eta-\lambda-2$. }

Combining~\eqref{eq:add12} and~\eqref{eq:mmm149}, we have
\begin{IEEEeqnarray}{rCl}
  \IEEEeqnarraymulticol{3}{l}{%
\E{c_1(X)+c_2(X)} 
}\nonumber\\*\,
&\leq&  -\log (1-\beta)  +\lambda\underbrace{\E{\log\left(1+\frac{X}{\lambda}\right)}}_{\leq \log\left(1+\frac{\mathcal{E}}{\lambda}\right)} +\left(1+\log\frac{1}{p}\right)\left(1+\frac{\text{Poi}_{\eta-2}(\eta-1)}{\eta-\lambda-2}\lambda\right)\underbrace{\E{X}}_{\leq \mathcal{E}} \nonumber\\
&&\>+\lambda\left(1+\log\frac{1}{p}\right)
  \text{Poi}_{\lambda}(\eta-1) \nonumber\\
&&\>+ \E{X\log\left(1+\frac{X}{\lambda}\right)\sum_{y=0}^{\eta-1}\text{Poi}_{\lambda+X}(y)-\log\beta\sum_{y=\eta}^{\infty}
  \text{Poi}_{\lambda+X}(y)} 
\\
&\leq&  -\log (1-\beta) + \lambda\log\left(1+\frac{\mathcal{E}}{\lambda}\right)+\left(1+\log\frac{1}{p}\right)\left(1+\frac{\text{Poi}_{\eta-2}(\eta-1)}{\eta-\lambda-2}\lambda\right)\mathcal{E}  \nonumber\\
&&+\lambda\left(1+\log\frac{1}{p}\right)
  \text{Poi}_{\lambda}(\eta-1) \nonumber\\
&&\>+ \mathsf{E} \Bigg[\underbrace{X\log\left(1+\frac{X}{\lambda}\right)\sum_{y=0}^{\eta-1}\text{Poi}_{\lambda+X}(y)-\log\beta\sum_{y=\eta}^{\infty}
  \text{Poi}_{\lambda+X}(y)}_{c_3(X)} \Bigg],
\label{eq:mmmm123}
\end{IEEEeqnarray}
where~\eqref{eq:mmmm123} follows from the concavity of $\log(\cdot)$ function. Now we bound the last term $c_3(X)$ at the RHS of~\eqref{eq:mmmm123}. By the law of total expectation,
\begin{IEEEeqnarray}{rCl}
 % \IEEEeqnarraymulticol{3}{l}{%
\E{c_3(X)}
%}\nonumber\\
&=&\Econd{c_3(X)}{X\leq \eta-\lambda}\textnormal{Pr}(X\leq \eta-\lambda)+\Econd{c_3(X)}{X > \eta-\lambda}\textnormal{Pr}(X > \eta+\lambda). 
\nonumber\\
\label{eq:cond1}
\end{IEEEeqnarray}
\llg{
For the first term at the RHS of~\eqref{eq:cond1}, notice that $\log \beta=-(\eta-\lambda)\log\frac{\eta}{\lambda}$, then
\begin{IEEEeqnarray}{rCl}
  \IEEEeqnarraymulticol{3}{l}{%
\Econd{c_3(X)}{X\leq \eta-\lambda} 
}\nonumber\\
&=&\mathsf{E}\Bigg[X\underbrace{\log\left(1+\frac{X}{\lambda}\right)}_{\leq \log \left(1+\frac{\eta-\lambda}{\eta}\right)}\sum_{y=0}^{\eta-1}
  \text{Poi}_{\lambda+X}(y)+(\eta-\lambda)\log\frac{\eta}{\lambda}\sum_{y=\eta}^{\infty}
  \text{Poi}_{\lambda+X}(y) \Bigg|X\leq \eta-\lambda \Bigg] \\
&\leq&\Econd{X\log\left(1+\frac{\eta-\lambda}{\lambda}\right)\sum_{y=0}^{\eta-1}
  \text{Poi}_{\lambda+X}(y)+(\eta-\lambda)\log\frac{\eta}{\lambda}\sum_{y=\eta}^{\infty}
  \text{Poi}_{\lambda+X}(y)}{X\leq \eta-\lambda} \nonumber \\
\\
&=&\mathsf{E} \Bigg[X\log\frac{\eta}{\lambda}\underbrace{\sum_{y=0}^{\eta-1}
  \text{Poi}_{\lambda+X}(y)}_{=1-\sum_{y=\eta}^{\infty}
  \text{Poi}_{\lambda+X}(y)}+(\eta-\lambda)\log\frac{\eta}{\lambda}\sum_{y=\eta}^{\infty}
  \text{Poi}_{\lambda+X}(y)\Bigg| X\leq \eta-\lambda \Bigg] \\
&=& \Econd{X\log\frac{\eta}{\lambda}+\log\frac{\eta}{\lambda}\sum_{y=\eta}^{\infty}(\eta-\lambda-X)
  \text{Poi}_{\lambda+X}(y)}{X\leq \eta-\lambda} 
\\
&=& \Econd{X\log\frac{\eta}{\lambda}}{X\leq \eta-\lambda} 
+
 \log\frac{\eta}{\lambda}\Econd{\sum_{y=\eta}^{\infty}(\eta-\lambda-X)
  \text{Poi}_{\lambda+X}(y)}{X\leq \eta-\lambda}. 
%\nonumber\\
\label{eq:ineq12}
\end{IEEEeqnarray}
The second term at the RHS of~\eqref{eq:ineq12} can be bounded as
\begin{IEEEeqnarray}{rCl}
  \IEEEeqnarraymulticol{3}{l}{%
\log\frac{\eta}{\lambda}\Econd{\sum_{y=\eta}^{\infty}(\eta-\lambda-X)
  \text{Poi}_{\lambda+X}(y)}{X\leq \eta-\lambda}
}\nonumber\\*\,
%&\leq& 2
%\log\frac{\eta}{\lambda}\Econd{\sum_{y=\eta}^{\infty}(\eta-\lambda-X)
%  \text{Poi}_{\lambda+X}(y)}{X\leq \eta-\lambda} \\
&=& 
\log\frac{\eta}{\lambda}\mathsf{E}\Bigg[\eta\sum_{y=\eta}^{\infty}
  \text{Poi}_{\lambda+X}(y) - \sum_{y=\eta}^{\infty}(\underbrace{\lambda+X}_{\leq \eta})
  \text{Poi}_{\lambda+X}(y)\Bigg |  X\leq \eta-\lambda \bigg] \\
&\leq& 
\log\frac{\eta}{\lambda}\Econd{\eta\sum_{y=\eta}^{\infty}
  \text{Poi}_{\lambda+X}(y) - \eta\sum_{y=\eta+1}^{\infty}
  \text{Poi}_{\lambda+X}(y)}{X\leq \eta-\lambda} \\
&=& 
\log\frac{\eta}{\lambda}\Econd{\eta\textnormal{Poi}_{\lambda+X}(\eta)}{X\leq \eta-\lambda} \\
&=& 
\log\frac{\eta}{\lambda}\Econd{\frac{e^{-(\lambda+X)}(\lambda+X)^{\eta}}{(\eta-1)!}}{X\leq \eta-\lambda} \\
&\leq& 
\log\frac{\eta}{\lambda}\Econd{\frac{e^{-\lambda}\lambda^{\eta}}{(\eta-1)!}+\frac{e^{-(\eta-1)}(\eta-1)^{\eta}}{(\eta-\lambda-1)(\eta-1)!}X}{X\leq \eta-\lambda}
\label{eq:firref2} \\
&\leq& 
\log\frac{\eta}{\lambda}\Econd{\frac{e^{-\lambda}\lambda^{\eta}}{(\eta-1)!}+\sqrt{\frac{\eta-1}{2\pi}}\frac{X}{\eta-\lambda-1}}{X\leq \eta-\lambda}
\label{eq:firref} \\
&=& 
\log\frac{\eta}{\lambda}\left(\frac{e^{-\lambda}\lambda^{\eta}}{(\eta-1)!}+\sqrt{\frac{\eta-1}{2\pi}}\frac{1}{\eta-\lambda-1}\Econd{X}{X\leq \eta-\lambda}\right),
\label{eq:mm122}
\end{IEEEeqnarray}
where~\eqref{eq:firref} follows from Stirling's bound: $(\eta-1)! \geq \sqrt{2\pi (\eta-1)} (\eta-1)^{\eta-1}e^{-(\eta-1)}$, and where~\eqref{eq:firref2} follows from the fact that when $X \leq \eta-\lambda$,
\begin{IEEEeqnarray}{rCl}
 e^{-(\lambda+X)}(\lambda+X)^{\eta}&=& e^{-\lambda}\lambda^{\eta}+\frac{e^{-(\lambda+X)}(\lambda+X)^{\eta} - e^{-\lambda}\lambda^{\eta}}{X}X \\
&\leq& e^{-\lambda}\lambda^{\eta}+\sup_{0 \leq X \leq \eta-\lambda}\left\{\frac{e^{-(\lambda+X)}(\lambda+X)^{\eta} }{X}\right\}X \\
&=& e^{-\lambda}\lambda^{\eta}+ \frac{e^{-(\eta-1)}(\eta-1)^{\eta}}{\eta-\lambda-1} X,
\end{IEEEeqnarray}
with the supremum being achieved at point $\eta-\lambda-1$. }

 Substituting~\eqref{eq:mm122} into~\eqref{eq:ineq12}, we have
\begin{IEEEeqnarray}{rCl}
  \IEEEeqnarraymulticol{3}{l}{%
\Econd{c_3(X)}{X\leq \eta-\lambda} 
}\nonumber\\
&\leq& \Econd{X\log\frac{\eta}{\lambda}}{X\leq \eta-\lambda}+
\log\frac{\eta}{\lambda}\left(\frac{e^{-\lambda}\lambda^{\eta}}{(\eta-1)!}+\sqrt{\frac{\eta-1}{2\pi}}\frac{1}{\eta-\lambda-1}\Econd{X}{X\leq \eta-\lambda}\right). \nonumber\\
\label{eq:sub11}
\end{IEEEeqnarray}

For the second term at the RHS of~\eqref{eq:cond1},
\begin{IEEEeqnarray}{rCl}
  \IEEEeqnarraymulticol{3}{l}{%
\Econd{c_3(X)}{X > \eta-\lambda} 
}\nonumber\\
&=&\mathsf{E}\Bigg[ X\log\left(1+\frac{X}{\lambda}\right)\sum_{y=0}^{\eta-1}
  \text{Poi}_{\lambda+X}(y)+(\eta-\lambda)\log\frac{\eta}{\lambda}\underbrace{\sum_{y=\eta}^{\infty}
  \text{Poi}_{\lambda+X}(y)}_{1-\sum_{y=0}^{\eta-1}
  \text{Poi}_{\lambda+X}(y)}\Bigg| X> \eta-\lambda \Bigg] 
\IEEEeqnarraynumspace\\
&=& \mathsf{E}\Bigg[(\underbrace{\eta-\lambda}_{\leq X})\log\frac{\eta}{\lambda}+\left(X\log\left(1+\frac{X}{\lambda}\right)-(\eta-\lambda)\log\frac{\eta}{\lambda}\right)\sum_{y=0}^{\eta-1}
  \text{Poi}_{\lambda+X}(y) \Bigg| X > \eta-\lambda \Bigg] \IEEEeqnarraynumspace
%&\leq&\Econd{X\log\left(1+\frac{X}{\lambda}\right)\sum_{y=0}^{\eta-1}
%  \text{Poi}_{\lambda+X}(y)+(\eta-\lambda)\log\frac{\eta}{\lambda}\sum_{y=\eta}^{\infty}
%  \text{Poi}_{\lambda+X}(y)}{X> \eta-\lambda} \\
\\
&\leq& \Econd{X\log\frac{\eta}{\lambda}}{X > \eta-\lambda} \nonumber\\
&&\>+
 \Econd{\left(X\log\left(1+\frac{X}{\lambda}\right)-(\eta-\lambda)\log\frac{\eta}{\lambda}\right)\sum_{y=0}^{\eta-1}
  \text{Poi}_{\lambda+X}(y)}{X > \eta-\lambda}. 
\label{eq:211}
\end{IEEEeqnarray}

The second term at the RHS of~\eqref{eq:211} can be bounded as
\begin{IEEEeqnarray}{rCl}
  \IEEEeqnarraymulticol{3}{l}{%
 \Econd{\left(X\log\left(1+\frac{X}{\lambda}\right)-(\eta-\lambda)\log\frac{\eta}{\lambda}\right)\sum_{y=0}^{\eta-1}
  \text{Poi}_{\lambda+X}(y)}{X > \eta-\lambda} 
}\nonumber\\*\,
&\leq&  \Econd{\sum_{y=0}^{\eta-1}\left(1+\log\left(1+\frac{X}{\lambda}\right)\right)(X-\eta+\lambda)
  \text{Poi}_{\lambda+X}(y)}{X > \eta-\lambda}  \label{eq:mmm324}\\
&\leq&  \mathsf{E}\Bigg[\left(1+\log\left(1+\frac{X}{\lambda}\right)\right)\sum_{y=0}^{\eta-1}\Bigg( \underbrace{(\lambda+X)
  \text{Poi}_{\lambda+X}(y)}_{= (y+1)\textnormal{Poi}_{\lambda+X}(y+1)} -\eta \sum_{y=0}^{\eta-1}
  \text{Poi}_{\lambda+X}(y) \Bigg) \Bigg| X > \eta-\lambda \Bigg] \nonumber\\
\\
&=&  \mathsf{E}\Bigg[\left(1+\log\left(1+\frac{X}{\lambda}\right)\right)\sum_{y=0}^{\eta-1}\Bigg(  {(\underbrace{y+1}_{_{\leq \eta}})\textnormal{Poi}_{\lambda+X}({y+1})} -\eta \sum_{y=0}^{\eta-1}
  \text{Poi}_{\lambda+X}(y) \Bigg) \Bigg| X > \eta-\lambda \Bigg] \nonumber\\
\\
&\leq&\Econd{\left(1+\log\left(1+\frac{X}{\lambda}\right)\right)\left(\eta\sum_{y=1}^{\eta}
  \text{Poi}_{\lambda+X}(y) -\eta \sum_{y=0}^{\eta-1}
  \text{Poi}_{\lambda+X}(y)\right)}{X > \eta-\lambda} \\
&=&\mathsf{E}\Bigg[\left(1+\log\left(1+\frac{X}{\lambda}\right)\right)\bigg( \eta
  \text{Poi}_{\lambda+X}(\eta) - \underbrace{\eta 
  \text{Poi}_{\lambda+X}(0)}_{\geq 0}\bigg)  \Bigg| X > \eta-\lambda \Bigg] \\
&\leq&  \Econd{\left(1+\log\left(1+\frac{X}{\lambda}\right)\right)\frac{e^{-(\lambda+X)}(\lambda+X)^{\eta}}{(\eta-1)!}}{X > \eta-\lambda} \\
&\leq& \sup_{X>\eta-\lambda} \left\{ \left(1+\log\left(1+\frac{X}{\lambda}\right)\right)\frac{e^{-(\lambda+X)}(\lambda+X)^{\eta}}{(\eta-1)!} \right\} 
\label{eq:supachi}\\
&=&\left(1+\log\frac{\eta+1}{\lambda}\right)\frac{e^{-(\eta+1)}(\eta+1)^{\eta}}{(\eta-1)!} 
\label{eq:m34d}\\
&\leq&\left(1+\log\frac{\eta+1}{\lambda}\right) \frac{\eta+1}{\sqrt{2\pi(\eta-1)}e^2}\left(\frac{\eta+1}{\eta-1}\right)^{\eta-1}.
\label{eq:sub12}
\end{IEEEeqnarray}
\llg{Here, Eq.~\eqref{eq:mmm324} is derived by applying the mean value theorem to the function $g(\xi)=\xi\log\left(1+\frac{\xi}{\lambda}\right)$, $\xi \geq \eta-\lambda$:
\begin{IEEEeqnarray}{rCl}
g(X)-g(\eta-\lambda) &=& g'(t) (X-\eta+\lambda),~t \in (\eta-\lambda, X) \\
&=& \left(1+\log\left(1+\frac{t}{\lambda}\right)\right) (X-\eta+\lambda),~t \in (\eta-\lambda, X) \\*
&\leq& \left(1+\log\left(1+\frac{X}{\lambda}\right)\right) (X-\eta+\lambda).
\end{IEEEeqnarray}
Here, Eq.~\eqref{eq:m34d} follows by the supremum in~\eqref{eq:supachi} being achieved at point $\eta-\lambda+1$, and~\eqref{eq:sub12} by the Stirling's bound: $(\eta-1)! \geq \sqrt{2\pi (\eta-1)} (\eta-1)^{\eta-1}e^{-(\eta-1)} $. 

\llg{Substituting}~\eqref{eq:sub12} into~\eqref{eq:211}, we have
\begin{IEEEeqnarray}{rCl}
%  \IEEEeqnarraymulticol{3}{l}{%
\Econd{c_3(X)}{X > \eta-\lambda} 
%}\nonumber\\
&\leq& \Econd{X\log\frac{\eta}{\lambda}}{X > \eta-\lambda}\nonumber\\
&&+ \left(1+\log\frac{\eta+1}{\lambda}\right) \frac{\eta+1}{\sqrt{2\pi(\eta-1)}e^2}\left(\frac{\eta+1}{\eta-1}\right)^{\eta-1}.
%  \nonumber\\
\label{eq:sub113}
\end{IEEEeqnarray}
}
\llg{Further substituting}~\eqref{eq:sub11} and~\eqref{eq:sub113} into~\eqref{eq:cond1}, we obtain
\begin{IEEEeqnarray}{rCl}
  \IEEEeqnarraymulticol{3}{l}{%
\E{c_3(X)}
}\nonumber\\
&\leq&\underbrace{\Econd{X\log\frac{\eta}{\lambda}}{X\leq \eta-\lambda}\textnormal{Pr}(X\leq \eta-\lambda)+ \Econd{X\log\frac{\eta}{\lambda}}{X > \eta-\lambda}\textnormal{Pr}(X > \eta-\lambda)}_{=\E{X\log\frac{\eta}{\lambda}}\leq \mathcal{E}\log\frac{\eta}{\lambda}}
\nonumber\\
&&\>+\log\frac{\eta}{\lambda}\left(\frac{e^{-\lambda}\lambda^{\eta}}{(\eta-1)!}+\sqrt{\frac{\eta-1}{2\pi}}\frac{1}{\eta-\lambda-1}\Econd{X}{X\leq \eta-\lambda}\right)\textnormal{Pr}(X\leq \eta-\lambda) \nonumber\\
&&\> + \left(1+\log\frac{\eta+1}{\lambda}\right) \frac{\eta+1}{\sqrt{2\pi(\eta-1)}e^2}\left(\frac{\eta+1}{\eta-1}\right)^{\eta-1}\textnormal{Pr}(X > \eta-\lambda) 
%&&\hspace{3.2cm}+ 2\log\frac{\eta+1}{\lambda}\frac{e^{-(\eta+1)}(\eta+1)^{\eta+1}}{(\eta-1)!} \textnormal{Pr}(X > \eta-\lambda)
\label{eq:mdd1344}
%\\
\\
&\leq& \mathcal{E}\log\frac{\eta}{\lambda} +\log\frac{\eta}{\lambda}\frac{e^{-\lambda}\lambda^{\eta}}{(\eta-1)!}\underbrace{\textnormal{Pr}(X\leq \eta-\lambda)}_{\leq 1} \nonumber
\\
&&\>+\log\frac{\eta}{\lambda}\sqrt{\frac{\eta-1}{2\pi}}\frac{1}{\eta-\lambda-1}\underbrace{\Econd{X}{X\leq \eta-\lambda}\textnormal{Pr}(X\leq \eta-\lambda)}_{\leq \E{X} \leq \mathcal{E}} \nonumber\\
&&\> + \left(1+\log\frac{\eta+1}{\lambda}\right) \frac{\eta+1}{\sqrt{2\pi(\eta-1)}e^2}\left(\frac{\eta+1}{\eta-1}\right)^{\eta-1}\textnormal{Pr}(X > \eta-\lambda) 
\\
&\leq& \mathcal{E}\log\frac{\eta}{\lambda} + \log\frac{\eta}{\lambda}\frac{e^{-\lambda}\lambda^{\eta}}{(\eta-1)!}+\log\frac{\eta}{\lambda}\sqrt{\frac{\eta-1}{2\pi}}\frac{\mathcal{E}}{\eta-\lambda-1} \nonumber\\
&&\> + \left(1+\log\frac{\eta+1}{\lambda}\right) \frac{\eta+1}{\sqrt{2\pi(\eta-1)}e^2}\left(\frac{\eta+1}{\eta-1}\right)^{\eta-1}\underbrace{\textnormal{Pr}(X > \eta-\lambda)}_{\leq \frac{\E{X}}{\eta-\lambda}\leq \frac{\mathcal{E}}{\eta-\lambda}} \\
&\leq& \mathcal{E}\log\frac{\eta}{\lambda} + \log\frac{\eta}{\lambda}\frac{e^{-\lambda}\lambda^{\eta}}{(\eta-1)!}+\log\frac{\eta}{\lambda}\sqrt{\frac{\eta-1}{2\pi}}\frac{\mathcal{E}}{\eta-\lambda-1} \nonumber\\
&&\> + \left(1+\log\frac{\eta+1}{\lambda}\right) \frac{\eta+1}{\sqrt{2\pi(\eta-1)}e^2}\left(\frac{\eta+1}{\eta-1}\right)^{\eta-1} \frac{\mathcal{E}}{\eta-\lambda},
%\nonumber\\
\label{eq:mdd134}
%+\Econd{c_1(X)
%+c_2(X)}{\textnormal{Pr}(X\leq \eta-\lambda)}\textnormal{Pr}(X > \eta+\lambda) \nonumber\\
\end{IEEEeqnarray}
where~\eqref{eq:mdd1344} follows by the law of total expectation, and~\eqref{eq:mdd134} by Markov's inequality. 

\llg{Before analyzing the asymptotics of each term at the RHS of~\eqref{eq:mdd134}, we first list some useful asymptotic results on some functions of $\eta$. 
%The proof is shown in Appendix~\ref{asympfuncb}. 
\begin{lemma}
\label{lem5}
Recalling $\eta$ denotes the largest integer that is less than or equal to the unique solution to 
\begin{IEEEeqnarray}{rCl}
(\eta-\lambda)\log\frac{\eta}{\lambda}=a_{\textnormal{P}} \log\frac{1}{\mathcal{E}}.
\label{eq:etaeq}
\end{IEEEeqnarray} 
Then,
\begin{IEEEeqnarray}{rCl}
\log \eta ~&\dot{=}&~ \log\log\frac{1}{\mathcal{E}}, \label{eq:logeta} \\
\eta~&\dot{=}&~\frac{a_{\textnormal{P}}\log\frac{1}{\mathcal{E}}}{\log\log\frac{1}{\mathcal{E}}}, \label{eq:eta}\\
\eta^{\eta}  ~&\dot{\geq}&~ \frac{1}{\mathcal{E}^{a_{\textnormal{P}}}} {\eta}^{\lambda-1}\lambda^{\eta-\lambda} \frac{\min\{\lambda,1\}}{e}, \label{eq:172approx}\\
e^{\eta} ~&\dot{\leq}&~e^{\lambda}{\mathcal{E}}^{-\frac{a_{\textnormal{P}}}{\log\frac{\eta}{\lambda}}}. 
\label{eq:173approx}
%\log\eta
\end{IEEEeqnarray}
\end{lemma}
\begin{IEEEproof}
See Appendix~\ref{asympfuncb}. 
\end{IEEEproof}  }

Now we bound each term at the RHS of~\eqref{eq:mdd134}. The first term scales as
\begin{IEEEeqnarray}{rCl}
\mathcal{E}\log\frac{\eta}{\lambda} &&~\dot{=}~\mathcal{E}\log{\eta}~\dot{=}~ \mathcal{E}\log\log\frac{1}{\mathcal{E}},
\label{eq:firstterm}
\end{IEEEeqnarray}
where~\eqref{eq:firstterm} follows from~\eqref{eq:logeta}.

For the second term,
\begin{IEEEeqnarray}{rCl}
\log\frac{\eta}{\lambda}\frac{e^{-\lambda}\lambda^{\eta}}{(\eta-1)!} &=& \eta\log\frac{\eta}{\lambda}\frac{e^{-\lambda}\lambda^{\eta}}{\eta!} \\
&~\dot{=}~&\underbrace{\eta\log\frac{\eta}{\lambda}}_{ ~\dot{=}~a_{\textnormal{P}} \log\frac{1}{\mathcal{E}} } \frac{e^{-\lambda}{(\lambda e)}^{\eta}}{\sqrt{2\pi \eta}{\eta}^{\eta}}\label{eq:stirapprox}\\
 &\dot{=}&~ a_{\textnormal{P}} \log\frac{1}{\mathcal{E}}\frac{e^{-\lambda}}{\sqrt{2\pi\eta}} \underbrace{\frac{{\lambda}^{\eta}}{\eta^{\eta}}}_{~\dot{\leq}~{\mathcal{E}^{a_{\textnormal{P}}}} \eta^{-\lambda+1}\lambda^{\lambda}e (\min\{\lambda,1\})^{-1}} \underbrace{e^{\eta}}_{~\dot{\leq}~e^{\lambda}{\mathcal{E}}^{-\frac{a_{\textnormal{P}}}{\log\frac{\eta+1}{\lambda}}}} \label{eq:186approx} \\
 &\dot{\leq}&~  \frac{a_{\textnormal{P}} \lambda^{\lambda}e}{\min\{\lambda,1\}\sqrt{2\pi\eta}\eta^{\lambda-1}} \mathcal{E}^{a_{\textnormal{P}}\left(1-\frac{1}{\log\frac{\eta+1}{\lambda}}\right)}  \log\frac{1}{\mathcal{E}}, 
\label{eq:secondterm}
\end{IEEEeqnarray}
where~\eqref{eq:stirapprox} follows by Stirling's approximation: $\eta!~\dot{=}~\sqrt{2\pi\eta}\left(\frac{\eta}{e}\right)^\eta$, and~\eqref{eq:186approx} by~\eqref{eq:172approx} and~\eqref{eq:173approx}.

The third term scales as
\begin{IEEEeqnarray}{rCl}
\underbrace{\log\frac{\eta}{\lambda}}_{~\dot{=}~\log\eta} \underbrace{\sqrt{\frac{\eta-1}{2\pi}}\frac{\mathcal{E}}{\eta-\lambda-1} }_{\frac{\mathcal{E}}{\sqrt{2\pi \eta}}} &&~\dot{=}~\frac{\log\eta}{\sqrt{2\pi\eta}}\mathcal{E} \\
&&~\dot{=}~\frac{\mathcal{E}}{\sqrt{2\pi\log\frac{1}{\mathcal{E}}}}\left(\log\log\frac{1}{\mathcal{E}}\right)^{\frac{3}{2}}, 
\label{eq:thirdterm}
\end{IEEEeqnarray}
where~\eqref{eq:thirdterm} follows by~\eqref{eq:eta}, and the fourth term as
\begin{IEEEeqnarray}{rCl}
\underbrace{\left(1+\log\frac{\eta+1}{\lambda}\right)}_{\dot{=}~\log \eta} \underbrace{\frac{\eta+1}{\sqrt{2\pi(\eta-1)}e^2}}_{\dot{=}~\frac{\sqrt{\eta}}{\sqrt{2\pi}e^2}}  \left(\frac{\eta+1}{\eta-1}\right)^{\eta-1} \underbrace{\frac{\mathcal{E}}{\eta-\lambda}}_{\dot{=}~\frac{\mathcal{E}}{\eta}}
 &&~\dot{=}~ \frac{\log \eta}{\sqrt{2\pi \eta}e^2}\underbrace{\left(\left(1+\frac{2}{\eta-1}\right)^{\frac{\eta-1}{2}}\right)^2}_{\dot{=}~e^2}\mathcal{E}\nonumber \\
\\
 &&~\dot{=}~ \frac{\log \eta}{\sqrt{2\pi \eta}}\mathcal{E} \label{eq:obvious} \\
 &&~\dot{=}~ \frac{\mathcal{E}}{\sqrt{2\pi\log\frac{1}{\mathcal{E}}}}\left(\log\log\frac{1}{\mathcal{E}}\right)^{\frac{3}{2}},
\label{eq:fourterm}
\end{IEEEeqnarray}
where~\eqref{eq:obvious} follows by~$\lim_{x \rightarrow 0}(1+x)^{\frac{1}{x}}=e$, and~\eqref{eq:fourterm} by~\eqref{eq:logeta} and~\eqref{eq:eta}. 

\llg{Comparing~\eqref{eq:firstterm} with~\eqref{eq:thirdterm} and~\eqref{eq:fourterm}, the first term dominates the third and fourth terms for $\mathcal{E} \to 0^{+}$. For the second term, by~\eqref{eq:firstterm} and~\eqref{eq:secondterm}, the ratio between it and the first term is bounded by $\frac{a_{\textnormal{P}} \lambda^{\lambda}e}{\min\{\lambda,1\}\sqrt{2\pi\eta}\eta^{\lambda-1}}\mathcal{E}^{a_{\textnormal{P}}\left(1-\frac{1}{\log\frac{\eta}{\lambda}}\right)-1}  \frac{\log\frac{1}{\mathcal{E}}}{\log\log\frac{1}{\mathcal{E}}}$. Recall that $a_{\textnormal{P}}>1$, and note that $\log\frac{\eta}{\lambda}$ tends to infinity as $\mathcal{E}$, then when $\mathcal{E}$ is small enough, we obtain $a_{\textnormal{P}}\left(1-\frac{1}{\log\frac{\eta}{\lambda}}\right)-1 > 0$. Hence, this ratio tends to zero as $\mathcal{E} \to 0^{+}$. Then, the first term also dominates the second terms. By~\eqref{eq:mdd134}, we have
\begin{IEEEeqnarray}{rCl}
 \E{c_3(X)}  ~\dot{\leq}~ \mathcal{E}\log\log\frac{1}{\mathcal{E}}.
\label{eq:c3(x)}
\end{IEEEeqnarray} }
Substituting~\eqref{eq:c3(x)} into~\eqref{eq:mmmm123}, and then into~\eqref{eq:mas123123}, we obtain
 \begin{IEEEeqnarray}{rCl}
%\limsup_{\mathcal{E} \to 0^{+}^+} 
{\mathsf{C}_{\textnormal{P}}(\mathcal{E})} &&~\dot{\leq}~ -\log (1-\beta) + \lambda\log\left(1+\frac{\mathcal{E}}{\lambda}\right)+\left(1+\log\frac{1}{p}\right)\left(1+\frac{\text{Poi}_{\eta-2}(\eta-1)}{\eta-\lambda-2}\lambda\right)\mathcal{E}  \nonumber\\
&&\hspace{0.6cm}+\>\lambda\left(1+\log\frac{1}{p}\right)
  \text{Poi}_{\lambda}(\eta-1) +  {\mathcal{E}{\log \log \frac{1}{\mathcal{E}} }}.
\label{eq:mm111}
\end{IEEEeqnarray}

Now we analyze the asymptotics of terms at the RHS of~\eqref{eq:mm111}. For the first and second terms, we have
\begin{IEEEeqnarray}{rCl}
-\log (1-\beta)~ &\dot{=}&~\beta =\mathcal{E}^{a_{\textnormal{P}}},  \label{eq:1244} \\
 \lambda\log\left(1+\frac{\mathcal{E}}{\lambda}\right) &~\dot{=}~& \mathcal{E}.
\label{eq:1245}
\end{IEEEeqnarray}

The third term can be bounded as
\begin{IEEEeqnarray}{rCl}
\left(1+\log\frac{1}{p}\right)\bigg(1+ \underbrace{ \frac{\text{Poi}_{\eta-2}(\eta-1)}{\eta-\lambda-2}}_{\leq \frac{1}{\eta-\lambda-2}}  \lambda \bigg)\mathcal{E} &\leq& \left(1+\log\frac{1}{p}\right)\left(1+\frac{1}{\eta-\lambda-2}\lambda\right)\mathcal{E} \\
 &\dot{=}& \lambda\left(1+\log\frac{1}{p}\right) \frac{\mathcal{E}}{\eta} \\
 &\dot{=}& \lambda\left(1+\log\frac{1}{p}\right) \frac{\mathcal{E}\log\log\frac{1}{\mathcal{E}}}{a_{\textnormal{P}}\log\frac{1}{\mathcal{E}}},
\label{eq:198lll}
\end{IEEEeqnarray}
where~\eqref{eq:198lll} follows by~\eqref{eq:eta}.

The fourth term can be bounded as
\begin{IEEEeqnarray}{rCl}
\lambda\left(1+\log\frac{1}{p}\right) \text{Poi}_{\lambda}(\eta-1) &=& \lambda\left(1+\log\frac{1}{p}\right) \frac{e^{-\lambda}\lambda^{\eta-1}}{(\eta-1)!} \\ &\dot{=}& \left(1+\log\frac{1}{p}\right) \frac{e^{-\lambda} \lambda^{\eta}} {\sqrt{2\pi (\eta-1)}} \left(\frac{e}{\eta-1}\right)^{\eta-1}
\label{eq:stirlll2} \\
&\dot{=}& \left(1+\log\frac{1}{p}\right) \frac{e^{-\lambda} \lambda^{\eta}} {\sqrt{2\pi \eta}} \left(\frac{e}{\eta}\right)^{\eta-1}\left(1+\frac{1}{\eta-1}\right)^{\eta-1} \\
&\dot{=}& \left(1+\log\frac{1}{p}\right) e^{-\lambda} \lambda^{\eta}\sqrt{\frac{\eta}{2\pi}}  \left(\frac{e}{\eta}\right)^{\eta}  \\
&\dot{=}& \left(1+\log\frac{1}{p}\right) \sqrt{\frac{\eta}{2\pi}} \frac{\lambda^{\lambda}}{\eta^{\lambda-1}} \frac{\min\{\lambda,1\}}{e} \mathcal{E}^{a_{\textnormal{P}}\left(1-\frac{1}{\log\frac{\eta}{\lambda}}\right)}  \log\frac{1}{\mathcal{E}}  ,
\IEEEeqnarraynumspace
\label{eq:203mm}
\end{IEEEeqnarray}
where~\eqref{eq:stirlll2} follows by Stirling's approximation: $(\eta-1)!~\dot{=}~\sqrt{2\pi(\eta-1)}\left(\frac{\eta-1}{e}\right)^{\eta-1}$, and~\eqref{eq:203mm} by~\eqref{eq:172approx} and~\eqref{eq:173approx}.

Comparing~\eqref{eq:1244},~\eqref{eq:1245},~\eqref{eq:198lll}, and~\eqref{eq:203mm} with the last term at the RHS of~\eqref{eq:mm111}, the last term still dominates for $\mathcal{E} \to 0^{+}$. Hence, we have
\begin{IEEEeqnarray}{rCl}
 \mathsf{C}_{\textnormal{G}}  ~\dot{\leq}~ \mathcal{E}\log\log\frac{1}{\mathcal{E}}.
\end{IEEEeqnarray}
Eq.~\eqref{eq:111} is proved. 
\subsubsection{Proof of Eq.~\eqref{eq:122}}
\label{secteq122}
We first present a useful lemma that bounds the left and right tail probabilities of \llg{the} Poisson distribution. \llg{
\begin{lemma}
\label{lem1}
Consider a Poisson random variable $W$ with parameter $\rho$. Then, for any $\xi>\rho$,
\begin{IEEEeqnarray}{rCl}
\textnormal{Pr}(W \geq \xi) \leq e^{-\xi\log\frac{\xi}{\rho}+\xi-\rho};
\end{IEEEeqnarray}
For any $\xi<\rho$,  
\begin{IEEEeqnarray}{rCl}
\textnormal{Pr}(W \leq \xi) \leq e^{-\xi\log\frac{\xi}{\rho}+\xi-\rho}. \label{eq:secinq}
\end{IEEEeqnarray}
\end{lemma}
\begin{IEEEproof}
See Appendix~\ref{C}.
\end{IEEEproof} }
%\begin{IEEEproof}
%See Appendix~\ref{C}.
%\end{IEEEProof}

Now we prove~\eqref{eq:122}. Consider a binary input $X_\text{B}$ with the distribution
\begin{IEEEeqnarray}{rCl}
p_{X_\text{B}}=\begin{cases}
 1-\frac{\mathcal{E}}{\eta_0} &\text{  if  } X_\text{B}=0, \\
 \frac{\mathcal{E}}{\eta_0} &\text{  if } X_\text{B}=\eta_0,
\end{cases}
\end{IEEEeqnarray}
where $\eta_0$ is the unique solution to 
\begin{IEEEeqnarray}{rCl}
\eta_0\log\frac{\eta_0}{\lambda}=a_{\textnormal{P}}\log\frac{1}{\mathcal{E}},
\label{eq:eta0value}
\end{IEEEeqnarray}
with $a_{\textnormal{P}}>1$. 

Given $Y$, denote $\hat{X}_B$ as the estimate of $X_\text{B}$ by the MAP decision rule~\eqref{eq:maprule}. Then the error probability $\text{P}_{\text{e}}$ by the MAP rule can be calculated as 
\begin{IEEEeqnarray}{rCl}
\text{P}_{\text{e}} &=& \textnormal{Pr}(X_\text{B}=0)\textnormal{Pr}\left(Y>\eta\right)+\textnormal{Pr}(X_\text{B}=x_0)\textnormal{Pr}\left(Y \leq \eta\right) \\
&=&\left(1-\frac{\mathcal{E}}{\eta_0}\right)\sum_{y=\eta}^{\infty}\textnormal{Poi}_{\lambda}(y) +  \frac{\mathcal{E}}{\eta_0}\sum_{y=0}^{\eta-1}\textnormal{Poi}_{\lambda+\eta_0}(y),\label{eq:peexp}
\end{IEEEeqnarray}
where 
\begin{IEEEeqnarray}{rCl}
\eta= \left\lfloor\frac{\eta_0+\log\frac{\eta_0-\mathcal{E}}{\mathcal{E}}}{\log\left(1+\frac{\eta_0}{\lambda}\right)}\right\rfloor,
\end{IEEEeqnarray}
 denotes the decision threshold of the likelihood ratio. For the convenience of calculation, we denote 
\begin{IEEEeqnarray}{rCl}
\eta\prime = \frac{\eta_0+\log\frac{\eta_0-\mathcal{E}}{\mathcal{E}}}{\log\left(1+\frac{\eta_0}{\lambda}\right)}.
\label{eq:etaprime22}
\end{IEEEeqnarray}
Note that $\eta= \left\lfloor \eta\prime \right\rfloor$, i.e., $\eta \leq \eta \prime < \eta+1$. The asymptotics of $\eta\prime$ can be shown as 
\begin{IEEEeqnarray}{rCl}
\eta\prime &=& \frac{\eta_0+\log\frac{\eta_0-\mathcal{E}}{\mathcal{E}}}{\log\left(1+\frac{\eta_0}{\lambda}\right)} \\
&=& \frac{\eta_0+\log\frac{1}{\mathcal{E}}+\log(\eta_0-\mathcal{E})}{\log\left(1+\frac{\eta_0}{\lambda}\right)}   \\
&=& \frac{\eta_0+\frac{1}{a_{\textnormal{P}}}\eta_0\log\frac{\eta_0}{\lambda}+\log(\eta_0-\mathcal{E})}{\log\left(1+\frac{\eta_0}{\lambda}\right)} \label{eq:198}  \\
&~\dot{=}~& \frac{\frac{1}{a_{\textnormal{P}}}\eta_0\log\frac{\eta_0}{\lambda}}{\log\left(1+\frac{\eta_0}{\lambda}\right)}   \\
&~\dot{=}~&\frac{\eta_0}{a_{\textnormal{P}}},
\label{eq:eta001}
\end{IEEEeqnarray}
where~\eqref{eq:198} follows from~\eqref{eq:eta0value}.

For the first term at the RHS of~\eqref{eq:peexp}, by Lemma~\ref{lem1},
\begin{IEEEeqnarray}{rCl}
\left(1-\frac{\mathcal{E}}{\eta_0}\right)\sum_{y=\eta}^{\infty}\textnormal{Poi}_{\lambda}(y) &\leq& \left(1-\frac{\mathcal{E}}{\eta_0}\right) e^{-\eta\log\frac{\eta}{\lambda}+\eta-\lambda} \\
% &=& \left(1-\frac{\mathcal{E}}{\eta_0}\right) e^{-\eta\log{\eta}+\eta(1+\log \lambda)-\lambda} \\
%&\leq& \left(1-\frac{\mathcal{E}}{\eta_0}\right) e^{-(\eta-1)\log\frac{\eta-1}{\lambda}+\eta-\lambda}\\
&\leq& \left(1-\frac{\mathcal{E}}{\eta_0}\right) e^{-(\eta\prime-1)\log\frac{\eta\prime-1}{\lambda}+\eta\prime-\lambda} \label{eq:exp12}\\
&=& \left(1-\frac{\mathcal{E}}{\eta_0}\right) e^{-(\eta\prime-1)\left(\log\left(1+\frac{\eta_0}{\lambda}\right)+\log\frac{\eta\prime-1}{\eta_0+\lambda}\right)+\eta\prime-\lambda} \label{eq:exp122}\\
&=& \left(1-\frac{\mathcal{E}}{\eta_0}\right) e^{-(\eta\prime-1)\log\left(1+\frac{\eta_0}{\lambda}\right)-(\eta\prime-1)\log\frac{\eta\prime-1}{\eta_0+\lambda}+\eta\prime-\lambda} \label{eq:exp122}\\
&=& \left(1-\frac{\mathcal{E}}{\eta_0}\right) e^{-(\eta\prime-1)\log\left(1+\frac{\eta_0}{\lambda}\right)-(\eta\prime-1)\log\frac{\eta\prime-1}{\eta_0+\lambda}+\eta\prime-\lambda} \label{eq:exp122}\\
&=& \underbrace{\left(1-\frac{\mathcal{E}}{\eta_0}\right)}_{\dot{=}~1} e^{-(\eta\prime-1)\log\left(1+\frac{\eta_0}{\lambda}\right)}\underbrace{e^{\left(1-\log\frac{\eta\prime-1}{\eta_0+\lambda}\right)\eta\prime}}_{~\dot{=}~e^{(1-\log\frac{1}{a_{\textnormal{P}}})\frac{\eta_0}{a_{\textnormal{P}}}}}\underbrace{e^{\log\frac{\eta\prime-1}{\eta_0+\lambda}-\lambda}}_{\dot{=}~e^{\log\frac{1}{a_{\textnormal{P}}}-\lambda}} 
\IEEEeqnarraynumspace\\
&\dot{=}&  e^{-(\eta\prime-1)\log\left(1+\frac{\eta_0}{\lambda}\right)} e^{(1-\log\frac{1}{a_{\textnormal{P}}})\frac{\eta_0}{a_{\textnormal{P}}}}e^{\log\frac{1}{a_{\textnormal{P}}}-\lambda} \label{eq:exp122}\\
&\dot{=}&  e^{-\left(\eta_0+\log\frac{1}{\mathcal{E}}+\log(\eta_0-\mathcal{E})-\log\left(1+\frac{\eta_0}{\lambda}\right)\right)} e^{(1-\log\frac{1}{a_{\textnormal{P}}})\frac{\eta_0}{a_{\textnormal{P}}}}e^{\log\frac{1}{a_{\textnormal{P}}}-\lambda} \label{eq:subeta} \\
&\dot{=}&  e^{(-1+\frac{1}{a_{\textnormal{P}}} + \frac{\log {a_{\textnormal{P}}}}{a_{\textnormal{P}}}){\eta_0}} \underbrace{e^{-\log\frac{1}{\mathcal{E}}}}_{=~\mathcal{E}} \underbrace{e^{-\left(\log(\eta_0-\mathcal{E})-\log\left(1+\frac{\eta_0}{\lambda}\right)\right)}}_{\dot{=}~1} e^{\log\frac{1}{a_{\textnormal{P}}}-\lambda} \IEEEeqnarraynumspace
\\
&\dot{=}& {\mathcal{E}} e^{(\frac{1+\log a_{\textnormal{P}}}{a_{\textnormal{P}}}-1)\eta_0-\log{a_{\textnormal{P}}}-\lambda} ,
%&\dot{=}& {\mathcal{E}} e^{(\frac{1+\log a_{\textnormal{P}}}{a_{\textnormal{P}}}-1)\eta_0} e^{-\log{a_{\textnormal{P}}}-\lambda}
\label{eq:mm333}
\end{IEEEeqnarray}
where~\eqref{eq:exp12} follows from $\eta\prime-1<\eta \leq \eta\prime$,~\eqref{eq:exp122} from~\eqref{eq:eta001}, and~\eqref{eq:subeta} from substituting~\eqref{eq:etaprime22} into~\eqref{eq:exp122}. 

Similarly, for the second term at the RHS of~\eqref{eq:peexp},
\begin{IEEEeqnarray}{rCl}
\frac{\mathcal{E}}{\eta_0}\sum_{y=0}^{\eta-1}\textnormal{Poi}_{\lambda+\eta_0}(y) &\leq& \frac{\mathcal{E}}{\eta_0} e^{-\eta\log\frac{\eta}{\lambda+\eta_0}+\eta-\eta_0-\lambda} \\
&\leq& \frac{\mathcal{E}}{\eta_0} e^{-(\eta\prime-1)\log\frac{\eta\prime-1}{\lambda+\eta_0}+\eta\prime-\eta_0-\lambda} \\
&=& \frac{\mathcal{E}}{\eta_0} \underbrace{e^{\left(1-\log\frac{\eta\prime-1}{\lambda+\eta_0}\right)\eta\prime-\eta_0}}_{\dot{=}~e^{(1-\log\frac{1}{a_{\textnormal{P}}})\frac{\eta_0}{a_{\textnormal{P}}}-\eta_{0}}} \underbrace{e^{\log\frac{\eta\prime-1}{\lambda+\eta_0}-\lambda}}_{\dot{=}~e^{\log\frac{1}{a_\textnormal{P}}-\lambda}} \\
&\dot{=}& \frac{{\mathcal{E}}}{\eta_0} e^{\left(\frac{1+\log a_{\textnormal{P}}}{a_{\textnormal{P}}}-1\right)\eta_0-\log{a_{\textnormal{P}}}-\lambda}.
\label{eq:mm444}
\end{IEEEeqnarray}
Substituting~\eqref{eq:mm333} and~\eqref{eq:mm444} into~\eqref{eq:peexp} yields
\begin{IEEEeqnarray}{rCl}
\text{P}_{\text{e}} 
&~\dot{\leq}~& {\mathcal{E}} e^{\left(\frac{1+\log a_{\textnormal{P}}}{a_{\textnormal{P}}}-1\right)\eta_0-\lambda} + \frac{{\mathcal{E}}}{\eta_0} e^{\left(\frac{1+\log a_{\textnormal{P}}}{a_{\textnormal{P}}}-1\right)\eta_0-\lambda} \\
&~\dot{=}~& {{\mathcal{E}}} e^{\left(\frac{1+\log a_{\textnormal{P}}}{a_{\textnormal{P}}}-1\right)\eta_0-\lambda}, \label{eq:finalpe}
\end{IEEEeqnarray} 
where~\eqref{eq:finalpe} follows from the fact that the first term dominates for $\mathcal{E} \to 0^{+}$, which can be shown by noting that $\eta_0 \rightarrow \infty$ for $\mathcal{E} \to 0^{+}$.

Since $X_\text{B}-Y-\hat{X}_\text{B}$ forms a Markov chain, following the same arguments as in~\eqref{eq:dpi}--\eqref{eq:fi}, we have
\begin{IEEEeqnarray}{rCl}
\mathsf{I}(X_\text{B};Y) 
% &\geq& \mathsf{I}(X_\text{B};\hat{X}_B) \label{eq:dpi} \\
%&=& \mathsf{H}(X_\text{B})-\mathsf{H}(X_\text{B}|\hat{X}_B) \\
&\geq& \mathsf{H}(X_\text{B})-\mathsf{H}_{\text{b}}(\text{P}_{\text{e}}).
\label{eq:fi22}
\end{IEEEeqnarray}
%where~$\mathsf{H}_{\text{b}}(\text{P}_{\text{e}}) \overset{\text{def}}{=}  -\text{P}_{\text{e}}\log \text{P}_{\text{e}} -(1-\text{P}_{\text{e}})\log (1-\text{P}_{\text{e}})$, and \eqref{eq:fi} follows by the Fano's inequality. 

Recalling  $\eta_0$ is the unique solution to \eqref{eq:eta0value}, and by using similar arguments as in~\eqref{eq:etaprime}--\eqref{eq:249}, we have
\begin{IEEEeqnarray}{rCl}
\log \eta_0 ~&\dot{=}&~ \log\log\frac{1}{\mathcal{E}} \label{eq:logeta0}, \\
\eta_{0}~ &\dot{=}& ~ \frac{a_{\textnormal{P}}\log\frac{1}{\mathcal{E}}}{\log\log\frac{1}{\mathcal{E}}}. \label{eta02}
\end{IEEEeqnarray}
 We bound $\mathsf{H}(X_\text{B})$ by 
\begin{IEEEeqnarray}{rCl}
\mathsf{H}(X_\text{B}) &=& -\frac{\mathcal{E}}{\eta_0}\log\frac{\mathcal{E}}{\eta_0}-\left(1-\frac{\mathcal{E}}{\eta_0}\right)\log\left(1-\frac{\mathcal{E}}{\eta_0}\right) \\
&\dot{=}& -\frac{\mathcal{E}}{\eta_0}\log\frac{\mathcal{E}}{\eta_0} \\
&\dot{=}& \frac{\mathcal{E}}{\eta_0}\left(\log\frac{1}{\mathcal{E}} +\log \eta_0 \right) \\
&\dot{=}& \frac{\mathcal{E}}{\eta_0}\log\frac{1}{\mathcal{E}} \label{eq:224}  \\
 &\dot{=}& \frac{1}{a_{\textnormal{P}}} \mathcal{E}{\log\log\frac{1}{\mathcal{E}}},
\label{eq:EEEE200}
\end{IEEEeqnarray}
where~\eqref{eq:224} follows from~\eqref{eq:logeta0}, and~\eqref{eq:EEEE200} from~\eqref{eta02}.

We bound $\mathsf{H}_{\text{b}}(\text{P}_{\text{e}})$ by
\begin{IEEEeqnarray}{rCl}
\mathsf{H}_{\text{b}}(\text{P}_{\text{e}}) &=&  -\text{P}_{\text{e}}\log \text{P}_{\text{e}} -(1-\text{P}_{\text{e}})\log (1-\text{P}_{\text{e}}) \\
%&\dot{=}& -\text{P}_{\text{e}}\log \text{P}_{\text{e}} + \text{P}_{\text{e}} -{\text{P}^2_e} \\
&\dot{=}& -\text{P}_{\text{e}}\log \text{P}_{\text{e}} \\
 &\dot{\leq}&  e^{\left(\frac{1+\log a_{\textnormal{P}}}{a_{\textnormal{P}}}-1\right)\eta_0-\lambda}{{\mathcal{E}}}\left(\log\frac{1}{\mathcal{E}} - \left(\frac{1+\log a_{\textnormal{P}}}{a_{\textnormal{P}}}-1\right)\eta_0 + \lambda \right) \label{eq:EEEE201}\\
 &\dot{=}&  e^{\left(\frac{1+\log a_{\textnormal{P}}}{a_{\textnormal{P}}}-1\right)\eta_0-\lambda}{{\mathcal{E}}}\log\frac{1}{\mathcal{E}},
\label{eq:EEEE20111}
\end{IEEEeqnarray}
where~\eqref{eq:EEEE201} follows by~\eqref{eq:finalpe}, and~\eqref{eq:EEEE20111} by the fact that the first term dominates, which can be shown by~\eqref{eta02}.

 Substituting~\eqref{eq:EEEE200} and~\eqref{eq:EEEE20111} into~\eqref{eq:fi22}, we obtain
\begin{IEEEeqnarray}{rCl}
\mathsf{I}(X_\text{B};Y) 
&~\dot{\geq}~& \frac{1}{a_{\textnormal{P}}} \mathcal{E}{\log\log\frac{1}{\mathcal{E}}} - e^{\left(\frac{1+\log a_{\textnormal{P}}}{a_{\textnormal{P}}}-1\right)\eta_0-\lambda}{{\mathcal{E}}}{\log\frac{1}{\mathcal{E}}} \label{eq:mmm2355} \\
&~\dot{=}~&\frac{1}{a_{\textnormal{P}}} \mathcal{E}{\log\log\frac{1}{\mathcal{E}}}, \label{eq:mmm235}
\end{IEEEeqnarray}
where~\eqref{eq:mmm235} follows by the fact that when $a_{\textnormal{P}}>1$,
\begin{IEEEeqnarray}{rCl}
\log a_{\textnormal{P}} = \log (1+a_{\textnormal{P}}-1) < a_{\textnormal{P}}-1,
\end{IEEEeqnarray}
 and by rarranging the terms, we get $\frac{1+\log a_{\textnormal{P}}}{a_{\textnormal{P}}}-1<0$. Then, $e^{\left(\frac{1+\log a_{\textnormal{P}}}{a_{\textnormal{P}}}-1\right)\eta_0-\lambda} \to 0$ as $\mathcal{E} \to 0^+$. Hence, the first term dominates.

Since $a_{\textnormal{P}}>1$ is chosen arbitrarily, 
\begin{IEEEeqnarray}{rCl}
\mathsf{C}_{\textnormal{P}} \geq \mathsf{I}(X_\text{B};Y)
~\dot{\geq}~ \sup_{a_{\textnormal{P}}>1} \frac{1}{a_{\textnormal{P}}} \mathcal{E}{\log\log\frac{1}{\mathcal{E}}}
&~\dot{=}~& \mathcal{E}{\log\log\frac{1}{\mathcal{E}}}.
\end{IEEEeqnarray}
Eq.~\eqref{eq:122} is proved.
%\end{IEEEproof}
\section{Conclusion}
\label{conclu}
This paper exactly \llg{characterizes} the low-SNR asymptotic capacity of two types of optical wireless channels when the inputs are subject to average-intensity constraints. The techniques used in this paper may be extended to analyze the low-SNR asymptotic capacity of the multiple-antenna optical wireless channels.
\section*{Acknowledgement}
\llg{The author would like to thank Dr. Ru-Han Chen and Dr. Jing Zhou for their fruitful discussions, which help improve the quality of the manuscript, and also the three anonymous reviewers for their valuable comments.}

\appendices
\llg{
\section{Proof of Lemma~\ref{lemm2}}
\label{lem:phifunc}

When $\tau=0$,~\eqref{eq:mg} obviously holds. When $\tau \geq \frac{\xi}{2}$, we have 
\begin{IEEEeqnarray}{rCl}
\phi(\xi)+\frac{2\tau}{\xi} \geq \phi(\xi)+\frac{2}{\xi}\cdot \frac{\xi}{2}=\phi(\xi)+1 > \frac{1}{\sqrt{2\pi}} \geq \phi(t-\xi).
\end{IEEEeqnarray} 
When $0 <\tau <\frac{\xi}{2}$, by the mean value theorem,
\begin{IEEEeqnarray}{rCl}
\frac{\phi(\xi)-\phi(\xi-\tau)}{\xi-(\xi-\tau)} =\phi'(\zeta),
\label{eq:rearrang}
\end{IEEEeqnarray}
where $\zeta$ is some point in the interval $(\xi-\tau,\xi)$, and $\phi'(\zeta)$ denotes the derivative of $\phi(\cdot)$ at point $\zeta$. Rearranging the terms in~\eqref{eq:rearrang}, we have
\begin{IEEEeqnarray}{rCl}
\phi(\xi-\tau) &=& \phi(\tau) - \phi'(\zeta)\tau \\
 &=& \phi(\tau) + \zeta\phi(\zeta)\tau \label{eq:deriva}\\
&=& \phi(\tau) + {\zeta}^2\phi(\zeta)\frac{\tau}{\zeta} \\
%&\leq& \phi(\tau) + \sup_{\zeta \in (\xi-\tau,\xi)}{\zeta}^2\phi(\zeta) \cdot \frac{\tau}{\zeta} \\
&\leq& \phi(\tau) + \frac{1}{\sqrt{2\pi}} 2e^{-1} \cdot \frac{\tau}{\zeta} \label{eq:supphi}\\
&\leq& \phi(\tau) + \frac{\tau}{\zeta} \\
&\leq& \phi(\tau) + \frac{\tau}{\xi-\tau} \\
&\leq& \phi(\tau) + \frac{\tau}{\xi-\frac{\xi}{2}} \\
&=& \phi(\tau) + \frac{2\tau}{\xi},
\end{IEEEeqnarray}
where~\eqref{eq:deriva} follows from $\phi'(\zeta)=-\zeta\phi(\zeta)$, and~\eqref{eq:supphi} follows from ${\zeta}^2\phi(\zeta)=\frac{1}{\sqrt{2\pi}}\zeta^2 e^{-\frac{\zeta^2}{2}} \leq \frac{1}{\sqrt{2\pi}} \sup_{z \in \mathbb{R}}\left\{  z^2 e^{-\frac{z^2}{2}}\right\} = \frac{1}{\sqrt{2\pi}}2e^{-1}$ with the supremum achieved at $z=\sqrt{2}$.
\section{Proof of Lemma~\ref{lem5}}
\label{asympfuncb}
 Denote $\eta\prime$ as the unique solution to~\eqref{eq:etaeq}, and then $\eta= \lfloor \eta\prime \rfloor$, i.e.,  $\eta \leq \eta\prime<\eta+1$. We have $\eta~\dot{=}~\eta\prime$ and $\log\eta~\dot{=}~\log\eta\prime$. 

Taking the logarithm at both sides of~\eqref{eq:etaeq}, we obtain 
\begin{IEEEeqnarray}{rCl}
\log (\eta\prime-\lambda)+\log\log\frac{\eta\prime}{\lambda} =\log\log\frac{1}{\mathcal{E}}-\log a_{\textnormal{P}}.
\label{eq:etaprime}
\end{IEEEeqnarray}
Since the first term at the RHS of~\eqref{eq:etaprime} dominates for $\mathcal{E} \to 0^{+}$,
 \begin{IEEEeqnarray}{rCl}
  \log (\eta\prime-\lambda)~\dot{=}~\log\log\frac{1}{\mathcal{E}}.
\end{IEEEeqnarray}
Then,~\eqref{eq:logeta} is proved by
\begin{IEEEeqnarray}{rCl}
\log\eta~\dot{=}~\log\eta\prime ~\dot{=}~  \log (\eta\prime-\lambda)~\dot{=}~\log\log\frac{1}{\mathcal{E}}.
\label{eq:246}
\end{IEEEeqnarray}
Dividing $\log\frac{\eta\prime}{\lambda}$ at both sides of~\eqref{eq:etaeq}, we obtain
\begin{IEEEeqnarray}{rCl}
\eta\prime -\lambda = \frac{a_{\textnormal{P}}\log\frac{1}{\mathcal{E}}}{\log\frac{\eta\prime}{\lambda}}.
\label{eq:etamm}
\end{IEEEeqnarray}
Then,~\eqref{eq:eta} is proved by
\begin{IEEEeqnarray}{rCl}
\eta~\dot{=}~\eta\prime -\lambda ~&\dot{=}&~ \frac{a_{\textnormal{P}}\log\frac{1}{\mathcal{E}}}{\log{\eta\prime}} \\
~&\dot{=}&~\frac{a_{\textnormal{P}}\log\frac{1}{\mathcal{E}}}{\log\log{\frac{1}{\mathcal{E}}}},
\label{eq:249}
\end{IEEEeqnarray}
where~\eqref{eq:249} follows from~\eqref{eq:246}. 

Removing the logarithm at both sides of~\eqref{eq:etaeq}, we obtain
\begin{IEEEeqnarray}{rCl}
\left(\frac{\eta\prime}{\lambda}\right)^{\eta\prime-\lambda}= \frac{1}{\mathcal{E}^{a_{\textnormal{P}}}}.
\label{eq:etaeta}
\end{IEEEeqnarray}
Rearranging the terms in~\eqref{eq:etaeta}, we have
\begin{IEEEeqnarray}{rCl}
{\eta\prime}^{\eta\prime} &=& \frac{1}{\mathcal{E}^{a_{\textnormal{P}}}} {\eta\prime}^{\lambda}\lambda^{\eta\prime-\lambda} \\
 &\geq& \frac{1}{\mathcal{E}^{a_{\textnormal{P}}}} {\eta}^{\lambda}\lambda^{\eta\prime-\lambda} \label{eq:comlam}\\
 ~&\dot{\geq}&~ \frac{1}{\mathcal{E}^{a_{\textnormal{P}}}} {\eta}^{\lambda}\lambda^{\eta-\lambda} \min\{\lambda,1\},
\label{eq:252}
\end{IEEEeqnarray}
where~\eqref{eq:comlam} follows form $\eta\prime \geq \eta$, and~\eqref{eq:252} follows from the fact that when $\lambda \leq 1$, $\lambda^{\eta\prime} \geq \lambda^{\eta+1}$, and when $\lambda > 1$, $\lambda^{\eta\prime} \geq \lambda^{\eta}$. Notice that 
\begin{IEEEeqnarray}{rCl}
\frac{ {\eta}^{\eta} }{{\eta\prime}^{\eta\prime}} &\geq& \frac{ {\eta}^{\eta} }{{(\eta+1)}^{\eta+1}} \\
&\geq& \frac{ {\eta}^{\eta+1} }{{(\eta+1)}^{\eta+1}} \frac{1}{\eta} \\
&=& \left( 1 - \frac{1}{\eta+1} \right)^{\eta+1} \frac{1}{\eta} \\
~&\dot{=}&~\frac{1}{\eta e}.
\label{eq:etaeta}
\end{IEEEeqnarray}
Combining~\eqref{eq:252} and~\eqref{eq:etaeta}, we obtain
\begin{IEEEeqnarray}{rCl}
\eta^{\eta}~&\dot{\geq}&~\frac{1}{\mathcal{E}^{a_{\textnormal{P}}}} {\eta}^{\lambda-1}\lambda^{\eta-\lambda} \frac{\min\{\lambda,1\}}{e}.
\end{IEEEeqnarray}
Eq.~\eqref{eq:172approx} is proved. 

By~\eqref{eq:etamm}, we have 
\begin{IEEEeqnarray}{rCl}
	e^{\eta\prime} &=& e^{\lambda}e^{\frac{a_{\textnormal{P}}\log\frac{1}{\mathcal{E}}}{\log\frac{\eta\prime}{\lambda}}} \\
&=& e^{\lambda}{\mathcal{E}}^{-\frac{a_{\textnormal{P}}}{\log\frac{\eta\prime}{\lambda}}} \\
~&\dot{\leq}&~ e^{\lambda}{\mathcal{E}}^{-\frac{a_{\textnormal{P}}}{\log\frac{\eta}{\lambda}}}.
\end{IEEEeqnarray}
Then, 
\begin{IEEEeqnarray}{rCl}
e^{\eta} \leq	e^{\eta\prime} 
~&\dot{\leq}&~ e^{\lambda}{\mathcal{E}}^{-\frac{a_{\textnormal{P}}}{\log\frac{\eta}{\lambda}}}.
\end{IEEEeqnarray}
Eq.~\eqref{eq:173approx} is proved. 
\section{Proof of Lemma~\ref{lem1}}
\label{C}

When $\xi > \rho$, for any $t>0$, we have 
\begin{IEEEeqnarray}{rCl}
\textnormal{Pr}(W \geq \xi) &=& \textnormal{Pr}(e^{tW} \geq e^{t\xi}) \\
&\leq& \frac{\E{e^{tW}}}{e^{t\xi}} \label{eq:chernoff} \\
&=& e^{\rho(e^t-1)-t\xi}, \label{eq:momentgenerate}
\end{IEEEeqnarray} 
where~\eqref{eq:chernoff} follows by the Chernoff bound, and~\eqref{eq:momentgenerate} by the fact that the moment generating function of the Poisson distribution is $\E{e^{tW}}=e^{\rho(e^t-1)}$. Since $\xi>\rho$, we let $t=\log\frac{\xi}{\rho} >0$, and the proof is concluded by substituting it into~\eqref{eq:momentgenerate}.

 We can prove~\eqref{eq:secinq} by using similar arguments but with $t$ being chosen negatively.
}
\bibliographystyle{IEEEtran}
\bibliography{./defshort1,./biblio1}

%\end{IEEEbiography}
%% insert where needed to balance the two columns on the last page with
%%\newpage


%\vfill


\end{document}